\documentclass[float,aps,prc,superscriptaddress,showpacs,twocolumn]{revtex4-1}
\usepackage{bm}
\usepackage{amsmath}
\usepackage{amssymb}
\usepackage{color}
\usepackage{epsfig}
\usepackage{hyperref}
\newcommand{\be}{\begin{equation}}
\newcommand{\ee}{\end{equation}}

\newcommand{\ber}{\begin{eqnarray}}
\newcommand{\eer}{\end{eqnarray}}

\newcommand{\bra}{\langle}
\newcommand{\ket}{\rangle}
\newcommand{\bs}[1]{\ensuremath{\boldsymbol{#1}}}

\begin{document}

\title {Giant and pigmy dipole resonances in $^{4}$He, $^{16,22}$O,
  and $^{40}$Ca from chiral nucleon-nucleon interactions}

\author {S. Bacca} \affiliation{TRIUMF, 4004 Wesbrook Mall, Vancouver,
  BC, V6T 2A3, Canada} \affiliation{Department of Physics and
  Astronomy, University of Manitoba, Winnipeg, MB, R3T 2N2, Canada}

\author{N. Barnea} \affiliation{Racah Institute of Physics, Hebrew
  University, 91904, Jerusalem}

\author{G. Hagen} \affiliation{Physics Division, Oak Ridge National
  Laboratory, Oak Ridge, TN 37831, USA} \affiliation{Department of
  Physics and Astronomy, University of Tennessee, Knoxville, TN 37996,
  USA}

\author{M. Miorelli} \affiliation{TRIUMF, 4004 Wesbrook Mall,
  Vancouver, BC, V6T 2A3, Canada} \affiliation{Department of Physics
  and Astronomy, University of British Columbia, Vancouver, BC, V6T
  1Z4, Canada}

\author{G.  Orlandini} \affiliation{Dipartimento di Fisica,
  Universit\`a di Trento, Via Sommarive 14, I-38123 Trento, Italy}
\affiliation{Istituto Nazionale di Fisica Nucleare, TIFPA, Via
  Sommarive 14, I-38123 Trento, Italy}

\author{T. Papenbrock} \affiliation{Department of Physics and
  Astronomy, University of Tennessee, Knoxville, TN 37996, USA}
\affiliation{Physics Division, Oak Ridge National Laboratory, Oak
  Ridge, TN 37831, USA}

\date{\today}

\begin{abstract}
  We combine the coupled-cluster method and the Lorentz integral
  transform for the computation of inelastic reactions into the
  continuum. We show that the bound--state--like equation
  characterizing the Lorentz integral transform method can be
  reformulated based on extensions of the coupled-cluster
  equation-of-motion method, and we discuss strategies for viable
  numerical solutions. Starting from a chiral nucleon-nucleon
  interaction at next-to-next-to-next-to-leading order, we compute the
  giant dipole resonances of $^4$He, $^{16,22}$O and $^{40}$Ca,
  truncating the coupled-cluster equation-of-motion method at the
  two-particle-two-hole excitation level. Within this scheme, we find
  a low-lying $E1$ strength in the neutron-rich $^{22}$O nucleus,
  which compares fairly well with data from [Leistenschneider {\it et
    al.} Phys. Rev. Lett. {\bf 86}, 5442 (2001)].  We also compute
 the electric dipole
  polariziability in $^{40}$Ca. Deficiencies of the employed
  Hamiltonian lead to overbinding, too small charge radii and a too
  small electric dipole polarizability in $^{40}$Ca.
\end{abstract}


\maketitle

\section{Introduction}

The inelastic response of an $A$-body system due to its interaction
with perturbative probes is a basic property in quantum physics. It
contains important information about the dynamical structure of the
system. For example, in the history of nuclear physics the study of
photonuclear reactions lead to the discovery of giant dipole
resonances (GDR)~\cite{BaK47}, originally interpreted as a collective
motion of protons against neutrons.  For neutron-rich nuclei far from the valley of
stability, such collective modes exhibit a 
fragmentation with low-lying strength, also called pigmy dipole
resonances (see, {\it e.g.}, Ref.~\cite{leistenschneider2001}),
typically interpreted as due to the oscillation of the excess neutrons
against a core made by all other nucleons.


Recently, progress was made in computing properties of medium mass and
some heavy nuclei from first-principles using a variety of methods
such as the coupled-cluster
method~\cite{kuemmel1978,bishop1991,hagen2013}, in-medium
similarity-renormalization-group
method~\cite{tsukiyama2011,hergert2013}, the self-consistent Green's
function method \cite{dickhoff2004,soma2013}, and lattice effective
field theory~\cite{laehde2014}. Although most of these methods focused
on bound-state properties of nuclei, there has been progress in
describing physics of unbound nuclear states and elastic
neutron/proton scattering with application to the neutron rich
helium~\cite{hagen2007d} and calcium
isotopes~\cite{hagen2012b,hagen2012c,hagen2013}. However, these
continuum calculations are currently limited to states that are of
single-particle like structure and below multi-nucleon breakup
thresholds.

The microscopic calculation of final--state continuum wave
functions of nuclei in the medium-mass regime constitutes still an
open theoretical problem.  This is due to the fact that at a given
continuum energy the wave function of the system has many different
components (channels) corresponding to all its partitions into
different fragments of various sizes.  Working in configuration space
one has to find the solution of the many--body Schr\"odinger equation
with the proper boundary conditions in all channels. The
implementation of the boundary conditions constitutes the main
obstacle to the practical solution of the problem.  In momentum space
the difficulty translates essentially into the proliferation with $A$
of the Yakubovsky equations as well as into the complicated treatment
of the poles of the resolvents. For example, the difficulties in
dealing with the three-body break-up channel for $^4$He have been
overcome only very recently~\cite{DeF12}.

The Lorentz integral transform (LIT) method~\cite{efros1994} allows to
avoid the complications of a continuum calculation, because it reduces
the difficulties to those typical of a bound--state problem, where the
boundary conditions are much easier to implement.  The LIT method has
been applied to systems with $A\leq 7$ using the Faddeev
method~\cite{martinelli1995}, the correlated hyperspherical-harmonics
method~\cite{efros1997a,efros1997d,efros1997c,efros1998,efros2000},
the EIHH method~\cite{bacca2002,bacca2004a,BaA04,gazit2006} or  the NCSM~\cite{stetcu2007,quaglioni2007}. All those methods,
however, have been introduced for dealing with typical few--body
systems and cannot be easily extended to medium--heavy nuclei.
Therefore it is desirable to formulate the LIT method in the framework
of other many--body methods. In the present work we present such a
formulation for the coupled--cluster (CC)
method~\cite{coester1958,coester1960,cizek1966,cizek1969,cizek1971,kuemmel1978},
which is a very efficient bound-state technique applied with success
on several medium--mass and a few heavy
nuclei~\cite{dean2004,kowalski2004,gour2006,hagen2008,hagen2010b,hagen2012a,hagen2012b,roth2012,binder2013,
  binder2014}, and see~\cite{hagen2013c} for a recent review. First
pioneering calculations of the GDR in $^{16}$O obtained by combining
the LIT with CC theory have been recently reported in a
Letter~\cite{bacca2013}.  In this paper, we will explain the details
of the approach and display comprehensive results on $^4$He, $^{16}$O,
and new results on the neutron-rich nucleus of $^{22}$O and on the
heavier nucleus of $^{40}$Ca.

The paper is organized as follows.  In Section \ref{sec:lit} a short
review of the LIT method is presented.  In Section \ref{sec:cclit} we
formulate it in the framework of CC theory and discuss two possible
strategies to solve the resulting equations. In Section~\ref{sec:He4}
we validate this method on $^4$He by benchmarking it against converged
EIHH calculations. In Sections~\ref{sec:O16}, \ref{sec:O22} and
\ref{sec:Ca40} we address the dipole response function of $^{16}$O,
$^{22}$O and $^{40}$Ca, respectively. Finally, in
Section~\ref{sec:conclusions} we draw our conclusions.

\section{The LIT method - a short review}
\label{sec:lit}

In order to determine cross sections due to external perturbative
probes one has to calculate various dynamical structure functions such
as
\begin{eqnarray} \label{response_ab}
&&S_{\alpha \beta}(\omega,q)=\\
\nonumber
&&\sum_{n} \langle 0
  |{\hat{\Theta}}^\dagger_{\alpha}(q) 
   |{n}\rangle\langle {n} |{\hat{\Theta}}_{\beta}(q)
   |0\rangle \delta (E_{ n} -E_0 -\omega )\:\mbox{.}
\end{eqnarray}
Here $\omega$ and $q$ are energy and momentum transfer of the external
probe, $|0\rangle$ and $| { n} \rangle$ denote ground and final state
wave functions of the considered system with energies $E_0$ and
$E_{n}$, respectively, while ${\hat{\Theta}}_{\alpha}$ denotes
excitation operators inducing transitions labeled by $\alpha$. The
$\sum_{ n}$ indicates both the sum over discrete state and an
integration over continuum Hamiltonian eigenstates.

For simplicity let us assume that 
${\hat{\Theta}}_{\alpha}(q)={\hat{\Theta}}_{\beta}(q)= {\hat{\Theta}} $
and consider the following inclusive structure function (also called response function)
\begin{equation} \label{response}
  S(\omega)=\sum_{ n} \langle 0
  |{\hat{\Theta}}^\dagger
   |{n} \rangle\langle {n} |{\hat{\Theta}}
   |0\rangle \delta (E_{n} -E_0 -\omega )\:\mbox{.}
\end{equation}
For few or many--body reactions with mass number $A > 2$ one very
often faces the problem that $S(\omega) $ cannot be calculated
exactly, since the microscopic calculation of $|{n}\rangle$ is too
complicated, due to the necessity to solve the many-body scattering
problem.  However, via the LIT approach the problem can be
reformulated in such a way that the knowledge of $|{n}\rangle$ is not
necessary \cite{efros1994}. To this end the integral transform of the
dynamical response function with a Lorentz kernel (LIT) is introduced
\begin{equation} \label{lorenzo}
  { L}(\omega_0,\Gamma )=\frac{\Gamma}{\pi}\int d\omega \frac{S(\omega)}{(\omega -\omega_0)
               ^2+\Gamma^2}\:\mbox{,}
\end{equation}   
where $\Gamma > 0 ${.}  The LIT method proceeds in two steps. First ${
  L}(\omega_0,\Gamma)$ is computed in a direct way, which does not
require the knowledge of $S(\omega)$, and then in a second step the
dynamical function is obtained from an inversion of the
LIT~\cite{efros1999,andreasi2005}.

The function ${ L}(\omega_0,\Gamma)$ can be calculated directly
starting from the definition in Eq.~(\ref{lorenzo}), substituting the
expression in (\ref{response}) for $S(\omega)$, and using the
completeness relation of the Hamiltonian eigenstates,
\begin{equation} \label{compl1}
   \sum_{ n} |{ n} \rangle \langle { n} | = 1 \:\mbox{.}
\end{equation}
Thus,
\begin{eqnarray}\label{lorenzog}
&&  { L}(\omega_0,\Gamma)=\frac{\Gamma}{\pi}\,\times\\
\nonumber
&&\langle 0 | {\hat{\Theta}}^{\dagger}\frac{1}{\hat
              {H}-E_0-\omega_0+i\Gamma}\frac{1}{\hat{H}-E_0-\omega_0-i\Gamma}\hat{\Theta}|0 \rangle \:\mbox{.}
\end{eqnarray}
The solutions $|\widetilde{\Psi}\ket$ of the equation
\begin{equation} \label{psi1}
  (\hat{H}-z )|\widetilde{\Psi}\rangle =\hat
   {\Theta} | 0\rangle 
\end{equation}
for different values of $\omega_0$ and $\Gamma$ lead directly to the
transform
\begin{equation} \label{elle}
  { L}(z)=\frac{\Gamma}{\pi}\bra \widetilde{\Psi} |\widetilde{\Psi}
  \rangle \:\mbox{.} 
\end{equation}
Here we introduced the quantity $z=E_0+\omega_0+i\Gamma$.  Since
${ L}(z)$ is finite the solution $|\widetilde{\Psi} \rangle$ of
Eq.~(\ref{psi1}) has the same asymptotic boundary conditions as a
bound--state.  Moreover the solution is unique. In fact if there were
two solutions $|\widetilde{\Psi}_1 \rangle$ and $|\widetilde{\Psi}_2
\rangle$, the hermiticity of $\hat{H}$, ensures that the homogeneous
equation
\begin{equation}
\label{psi1a}
(\hat{H}-z) (|\widetilde{\Psi}_1\rangle - |\widetilde{\Psi}_2 \rangle)=0
\end{equation}
has only the trivial solution $(|\widetilde{\Psi}_1\rangle -
|\widetilde{\Psi}_2 \rangle) =0$.

From the inversion of the calculated ${ L}(\omega_0,\Gamma)$ one
obtains the dynamical function $S(\omega)$.  The LIT method leads to
an exact response function as shown in benchmarks with other methods
for two- and three-body systems \cite{lapiana2000, golak2002}.  In the
reviews~\cite{efros2007b,bacca2014} the interested reader can find
more details on the LIT method, on the generalizations to exclusive
and hadronic processes as well as its application to various
electro-weak interactions with light nuclei.

\section{The LIT in Coupled--Cluster Theory}
\label{sec:cclit}

In coupled-cluster theory we work with the similarity transformed
Hamiltonian 
\begin{equation}\label{hbar}
 \overline{H} = \exp(-T) \hat H_{N} \exp(T) .
\end{equation}
Here $H_{N}$ is normal-ordered with respect to a chosen uncorrelated
reference state $|\Phi_0\rangle$, which is typically the Hartree-Fock
state. Correlations are introduced through the cluster operator $T$
which is a linear combination of particle-hole ($ph$) excitation
operators, i.e. $T=T_1+T_2+\ldots $, with the $1p$-$1h$ excitation
operator $T_1$, the $2p$-$2h$ excitation operator $T_2$, and so on.
The similarity transformed Hamiltonian~(\ref{hbar}) is non-Hermitian
and has left- and right eigenstates which constitute a complete
bi-orthogonal set according to 
\be \label{cc_complete} \bra
{n}_L|{n'}_R \ket = \delta_{n,n'}, \:\:\: \sum_{ n}|{ n}_R \ket\bra {
  n}_L|= 1 \,.  
\ee 
We note that the right ground-state is nothing but the reference
state, i.e. $|0_R\ket = |\Phi_0\ket$, while the corresponding left
ground-state is given by $\bra 0_L | = \bra \Phi_0 |(1+\Lambda)$. Here
$\Lambda$ is a linear combination of particle-hole de-excitation
operators, see e.g. \cite{bartlett2007}.

Using the left and right eigenstates we can define the response
function corresponding to the similarity-transformed Hamiltonian $\overline{H}$
analogous to Eq.~(\ref{response}) 
\begin{equation} \label{cc_response} S(\omega)=\sum_{n} \bra 0_L |
  \overline{ \Theta^\dagger} | { n}_R \ket \bra {n}_L |
  \overline{\Theta} | 0_R \ket \delta (E_{ n} -E_0 -\omega ).
\end{equation}
The similarity-transformed 
excitation operators 
\begin{equation} \label{theta_bar}
\overline \Theta = \exp(-T)\hat \Theta\exp(T), \:\:
\overline{\Theta^{\dagger}} = \exp(-T)\hat\Theta^{\dagger}\exp(T)
\end{equation}
enters in Eq.~(\ref{cc_response}).  The Baker-Campbell-Hausdorff
expansion of $\overline\Theta$ terminates exactly at doubly nested
commutators in the case of a one-body operator and at quadruply nested
commutators in the case of a two-body operator (see, for example,
\cite{shavittbartlett2009} for more details).  Substituting
Eq.~(\ref{cc_response}) in Eq.~(\ref{lorenzo}) and using the
completeness relation (\ref{cc_complete}), one obtains
\begin{equation}\label{cc_lorentz0bar}
  { L}(\omega_0,\Gamma)=\frac{\Gamma}{\pi}\,\langle 0_L | 
         {\overline{\Theta^{\dagger}}}\frac{1}{\overline
              {H}-z^*}\frac{1}{\overline{H}-z}\overline{\Theta} |0_R \rangle,
\end{equation}
with $z=E_0+\omega_0+i \Gamma$,
and in analogy with Eq.~(\ref{elle}) one has
\begin{equation}\label{psinorm}
 { L}(z)\equiv\frac{\Gamma}{\pi}\, \langle\widetilde{\Psi}_L(z^*)\vert\widetilde{\Psi}_R(z)\rangle\,.
\end{equation}
Here $\langle\widetilde{\Psi}_L\vert$ and 
$|\widetilde{\Psi}_R\rangle$ satisfy the equations
\begin{eqnarray}  
  (\overline{H}-z)|\widetilde{\Psi}_R(z)\rangle &=&
  \overline{\Theta} | 0_R\rangle \label{cc_psi1} \\
  \langle\widetilde{\Psi}_L(z^*)\vert(\overline{H}-z^*) &=&
  \langle 0_L\vert \overline{\Theta^\dagger}\,. \label{cc_psi2}
\end{eqnarray}
Since $L(z)$ is finite, $|\widetilde{\Psi}_R(z)\ket$ and $\bra
\widetilde{\Psi}_L(z^*)|$ must have bound-state like boundary conditions.

Eq.~(\ref{cc_lorentz0bar}) can also be written as
\begin{equation}\label{cc_lorentz2}
 { L}(z)=-\frac{1}{2 \pi} {\rm Im}\left\{
            \bra 0_L | {\overline{\Theta^{\dagger}}}
             \left(\frac{1}{\overline{H}-z^*} 
              - \frac{1}{\overline{H}-z}\right)
             \overline{\Theta} |0_R \rangle \right\},
\end{equation}
or as
\begin{equation} \label{cc_elle1}
    { L}(z )=-\frac{1}{2  \pi} {\rm Im}\left\{
        \bra \widetilde{\Psi}_L(z^*) \vert \overline{\Theta} 
         \vert 0_R\ket - 
         \bra 0_L | \overline{\Theta^\dagger} 
         |\widetilde{\Psi}_R(z)\rangle \right\}
   \,.
\end{equation}

One can also write $L(z)$ in function of $|\widetilde{\Psi}_R (z)\rangle $
only, or of $\bra \widetilde{\Psi}_L(z^*)|$ only, as
\begin{equation} \label{cc_elle2}
  { L}(z )=-\frac{1}{2  \pi} {\rm Im}\left\{
        \bra 0_L | \overline{\Theta^\dagger}\left(|\widetilde{\Psi}_R(z^*)\rangle-  
         |\widetilde{\Psi}_R(z)\rangle\right) 
       \right\}
\end{equation}
\begin{equation} \label{cc_elle3}
  { L}(z )=-\frac{1}{2  \pi} {\rm Im}\left\{
        \left(\bra \widetilde{\Psi}_L(z^*) \vert - \bra \widetilde{\Psi}_L(z)|\right) \overline{\Theta} 
         \vert 0_R\ket 
       \right\}
   \,.
\end{equation}

Within the coupled--cluster theory, equations (\ref{cc_psi1}) and
(\ref{cc_psi2}), to which we shall refer as the
Lorentz-Integral-Transform Coupled-Cluster (LIT-CC) equations, are
equivalent to Eq.~(\ref{psi1a}) introduced in the previous
Section. They are the key equations to solve to calculate $L(z)$ via
either Eq.~(\ref{psinorm}), or any of Eqs.
(\ref{cc_lorentz2})--(\ref{cc_elle3}). It is important to remark that
in deriving Eqs.~(\ref{psinorm}),
(\ref{cc_lorentz2})--(\ref{cc_elle3}) no approximation has been made.

\subsection{Solving the LIT-CC equations}
\label{sec:ccliteqs}

As we have seen in the previous Section, being able to solve
Eqs.~(\ref{cc_psi1}) and/or (\ref{cc_psi2}) to sufficient accuracy is the
key for the success of the method.  To solve the LIT-CC equations one
may proceed in a way analogous to what is done in the
equation-of-motion coupled-cluster method for excited states
~\cite{stanton1993}. We therefore write the wave function
$|\widetilde{\Psi}_R(z)\ket$ in the form,
\begin{eqnarray} \label{eom1}\nonumber
|\widetilde{\Psi}_R(z)\ket &=&{\cal R}(z) | \Phi_0\ket\equiv\left(r_0(z) + 
\sum_{i a} r_{i}^{{a}}(z) \hat{c}^\dagger_{{a}} \hat{c}_{{i}}+\right.\\
&+&\left. \frac{1}{4} \sum_{i j a  b} r_{i j}^{{a b}}(z) \hat{c}^\dagger_{a}\hat{c}^\dagger_{b}    
\hat{c}_{j}\hat{c}_{i} + \ldots \right)|\Phi_0\ket \;,
\end{eqnarray}
and analogously for
$\bra\widetilde{\Psi}_L(z^*)|$ in the form,
\begin{eqnarray} \label{eomL}\nonumber
\bra \widetilde{\Psi}_L(z^*)|&=& \bra \Phi_0 | {\mathcal L}(z^*)\equiv
\bra \Phi_0 | \left( l_0(z^*) + \sum_{i a} l_{i}^{{a}}(z^*)
                  \hat{c}^\dagger_{{i}} \hat{c}_{{a}}+\right.\\
               &+&\left.\frac{1}{4} \sum_{i j a  b} l_{i j}^{{a b}}(z^*) 
                  \hat{c}^\dagger_{i}\hat{c}^\dagger_{j} 
                  \hat{c}_{b}\hat{c}_{a}
               + \ldots
    \right) \;.
\end{eqnarray}
Substituting $|\widetilde{\Psi}_R(z)\ket$  in Eq.~(\ref{cc_psi1}) yields
\begin{equation} \label{calc_rr}
 (\overline{H}-z) {\cal R}(z)
 |{\Phi}_0\ket = \overline{\Theta} | 0_R \ket \;\;,
\end{equation}
and similarly for the left equation, 
\begin{equation} \label{calc_ll}
  \bra \Phi_0|{\cal L}(z^*)(\overline{H}-z^*)  = \bra 0_L |\overline{\Theta^\dagger}  \;\;.
\end{equation}
Projecting the last two equations on $n$-particle $n$-hole excited
reference states we get a set of linear equations for the amplitude
operators ${\mathcal R}(z)$ and ${\mathcal L}(z^*)$.  These equations
are similar to the CC equations-of-motion~\cite{stanton1993} up to the
source term on the right--hand--side. As
$\overline{H}$ is a scalar under rotatations, the amplitudes
${\mathcal R}(z)$ and ${\mathcal L}(z^*)$ exhibit the same
symmetries as the excitation operators $\overline\Theta$ and
$\overline{\Theta^\dagger}$, respectively. Once these equations are
solved one obtains the LIT as \be
\label{psir_psil}
 { L}(z) = \frac{\Gamma}{\pi} \langle \Phi_0 \vert {\mathcal
   L}(z^*){\mathcal R}(z)\vert \Phi_0\rangle\,, 
\ee
or\be
\label{cc_elle_r}
 {  L}(z)=-\frac{1}{2\pi} {\rm Im}\left\{
    \bra \Phi_0\vert {\mathcal L}(z^*) \overline{\Theta}\vert 0_R \ket - 
    \bra 0_L \vert \overline{\Theta^\dagger} 
    {\cal R}(z) \vert \Phi_0 \ket
    \right\} \,,
\ee
or
\begin{equation} \label{cc_elle_r1}
  { L}(z )=-\frac{1}{2\pi} {\rm Im}\left\{
        \bra 0_L| \overline{\Theta}^\dagger\left({\cal R}(z^*)-  
         {\cal R}(z)\right)\vert \Phi_0 \ket 
       \right\}\,,
       \end{equation}
       or
\be
\label{cc_elle_r2}
 {  L}(z)=-\frac{1}{2\pi} {\rm Im}\left\{
    \bra \Phi_0\vert ({\mathcal L}(z^*) -  {\mathcal L}(z) )\overline{\Theta}\vert 0_R \ket  
    \right\}\,. 
\ee

We note that $L(z)$ can be computed by solving
Eqs.~(\ref{calc_rr})-(\ref{calc_ll}) and any of the
Eqs.~(\ref{psir_psil})-(\ref{cc_elle_r2}). 
These equations provide different,
but equivalent, ways of obtaining the LIT. 
This gives us a valuable tool to check the
implamentation of the LIT-CC method. 
On test examples these different approaches gave identical numerical results.

To obtain $L(z)$ one is required
to solve the equations of motion (\ref{psir_psil})-(\ref{cc_elle_r2})
for every different value of $z$,
thus making it not very convenient, especially if the model space size
is large.  It is thus convenient to reformulate the solutions of these
equations by using the Lanczos algorithm.

\subsection{The Lanczos method}

Here we generalize the Lanczos approach of Ref.~\cite{efros2007} to
non-Hermitian operators and thereby avoid solving Eqs. (\ref{calc_rr})
and (\ref{calc_ll}) for every different value of $z$. Starting, {\it
  e.g.}, from Eq.~(\ref{cc_lorentz2}) one can write the $L(z)$ in
matrix form on the particle-hole basis as
\begin{eqnarray}\label{LITM}
{ L}(z)&=&-\frac{1}{2\pi} {\rm Im} \left[ {\bf S}^L \left((\bs{M}-z^*)^{-1}\right.\right.\\
\nonumber
&&
    \left.\left.-(\bs{M}-z)^{-1}\right) \bs{S}^R   \right]\,,
\end{eqnarray}
where the matrix elements $M_{\alpha,\alpha'}$ of $\bs{M}$ and the components $S^R_{\alpha}$ and $S^L_{\alpha}$ of the row- and column-vectors 
$\bs{S}^L$ and $\bs{S}^R$  are given by
\ber 
M_{\alpha,\alpha'}&=& \bra \Phi_{\alpha} | \overline H
 |\Phi_{\alpha'}\ket \,,\\
S^R_{\alpha}&=&\bra \Phi_\alpha |\overline \Theta | 0_R\ket \,,\\
S^L_{\alpha}&=&\bra 0_L|\overline{\Theta^\dagger} | \Phi_\alpha \ket\,. 
\eer
The indices $\alpha,\alpha'$ run over the
$0p$-$0h$,$1p$-$1h$,$2p$-$2h$,$\ldots $ states
\begin{equation}
 |\Phi_0\ket, |\Phi_i^a\ket=\hat c_a^{\dagger}\hat c_i |\Phi_0\ket
,|\Phi_{ij}^{ab}\ket
  =\hat c_a^{\dagger}\hat c_b^{\dagger} \hat c_i \hat c_j |\Phi_0\ket
  , \ldots \;.
\end{equation}
Notice that 
$\bs{S}^L  \bs{S}^R= \bra 0_L|\overline{\Theta^\dagger}\,
\overline \Theta | 0_R\ket=\bra 0|\Theta^\dagger\,\Theta | 0\ket$.

At this point we can make use of the Lanczos algorithm  to evaluate $L(z)$. However, since 
the matrix $\bs{M}$ is nonsymmetric, we use its complex symmetric variant~\cite{Col95}.
To this end we define the left and the right vectors 
\ber
  \bs{v}_0 &=& \bs{S}^R/\sqrt{\bs{S}^L \bs{S}^R } \cr
  \bs{w}_0 &=& \bs{S}^L/\sqrt{\bs{S}^L \bs{S}^R }\,.
\eer
Equation (\ref{LITM}) becomes
\begin{eqnarray}
&& {L}(z)=-\frac{1}{2\pi}\rm{Im}\left\{\left[\bs{S}^L\bs{S}^R\right]\right. \\
\nonumber
&&
     \left.  {\bs{w}_0}\left[(\bs{M}-z^*)^{-1}-(\bs{M}-z)^{-1}\right] \bs{u}_0 \right\}\,.
\end{eqnarray}

One notices that the LIT depends on the matrix element
\begin{equation}
  x_{00}\equiv \bs{w}_0     (\bs{M}-z)^{-1}  \bs{u}_0 \:.
\end{equation}
One can calculate $x_{00}$ applying Cramer's rule to the solution of the 
linear system 
\begin{equation}
\label{linsys0}
  \sum_{\beta} (\bs{M}-z)_{\alpha\beta}x_{\beta 0}=\delta_{\alpha 0}\,,
\end{equation}
which arises from the identity 
\begin{equation}
  (\bs{M}-z)(\bs{M}-z)^{-1}=I 
\end{equation}
on the Lanczos basis 
$\{\bs{u}_i,\bs{w}_i \mbox{,} i=0\mbox{,}\ldots\mbox{,} n-1\}$.
In the Lanczos basis $\bs{M}$ takes on a
tridiagonal form 
\begin{equation}\label{m00}
\bs{M}=\left(\begin{array}{ccccc}
a_0    &  b_{0}    & 0     &  0     & \cdots  \\
b_0    &  a_{1}    & b_{1}  &  0     & \cdots  \\
0      &  b_{1}    & a_{2}  &  b_{2} & \cdots  \\
0      &   0      & b_{2}  &  a_{3} & \cdots  \\ 
\vdots & \vdots   & \vdots & \vdots & \ddots
\end{array}\right)\:\mbox{.}
\end{equation}
In this way one is able to write $x_{00}$ as a continued fraction containing 
the Lanczos coefficients $a_i$ and $b_i$, 
\begin{equation} \label{conti}
  x_{00}(z)=\frac{1}{a_{0}-z+\frac{b^{2}_{1}}{a_{1}-z+
                  \frac{b^{2}_{2}}{a_{2}-z+b^{2}_{3}\ldots}}} \:,
\end{equation}
and thus also the LIT becomes a function of the Lanczos coefficients
\be
 { L}(z)=-\frac{1}{2\pi}{\rm Im}\left\{\left[\bs{S}^L   \bs{S}^R\right]
      \left[ x_{00}(z^*)-x_{00}(z) \right]\right\} \ .
\ee
This illustrates that the Lanczos method allows to determine $L(z)$ 
without inverting the Hamiltonian matrix.
\begin{figure}[ht]
\includegraphics[scale=0.4,clip=]{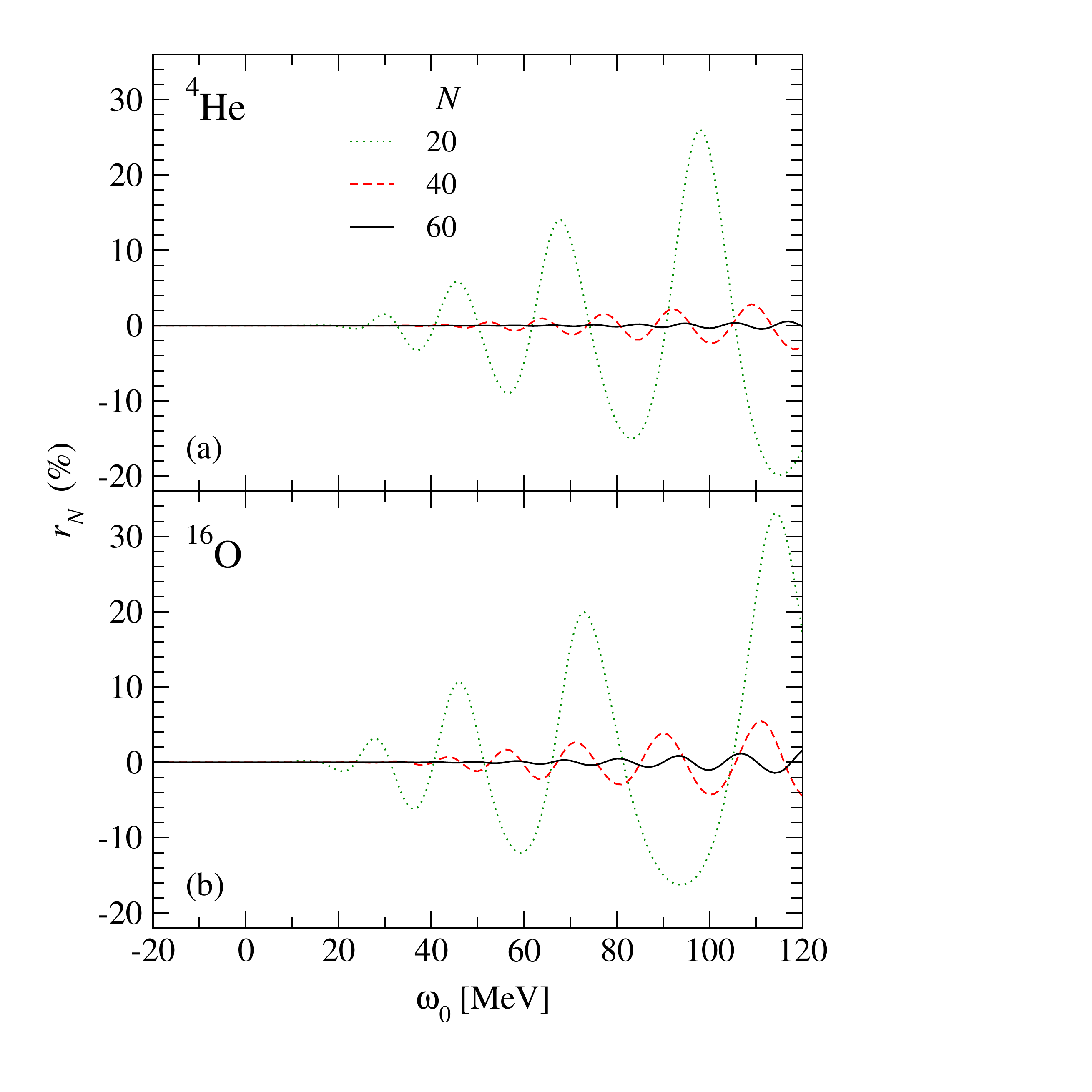}
\caption{(Color online) The relative convergence of the LIT 
  $L(\omega_0,\Gamma)$ as a function
  of the number $N$ of
  Lanczos steps: $^4$He in panel
  (a) and $^{16}$O in panel (b). The LIT are calculated for
  $\Gamma=10$ MeV.}
\label{fig_lancz_conv}
\end{figure}

The Lanczos approach outlined above has few important advantages for
the LIT method. First of all, the tridiagonalization of ${\bf M}$ has
to be done only once regardless of the value of $\omega_0$ and
$\Gamma$.  Moreover, one can usually converge with reasonably few
Lanczos vectors (depending on the nucleus and the excitation operator
$\Theta$).  This is expected since at low values of $\omega_0$ the LIT
is dominated by the lowest eigenvalues of ${\bf M}$, and for $\omega_0
\longrightarrow\infty$, $L(z)$ is dominated by the first Lanczos
vector.

Figure~\ref{fig_lancz_conv} shows the fast convergence rate (for a dipole operator) by
showing the ratio 
\be r_{N}(\omega_0, \Gamma)= [{
  L}_{N}(\omega_0,\Gamma) - { L}(\omega_0,\Gamma)]/ {
  L}(\omega_0,\Gamma) \times 100 \,, 
\ee 
where ${ L}_{N}(\omega_0,\Gamma)$ is the LIT calculated with ${N}$
Lanczos steps and ${ L}(\omega_0,\Gamma)$ is the converged result.
The curves are obtained with
$\Gamma=10$~MeV.
The convergence is indeed very fast and with 60 Lanczos steps one can
reach a numerical precision which is below one percent, and
about 90
Lanczos vectors are sufficient to reach convergence.

\subsection{Removal of spurious states}\label{sec:remcm}
  
In this paper, we will apply the LIT-CCSD method to the computation of
the dipole response and use the excitation operator $\Theta \equiv
\hat{\bf D}$ where $\hat{\bf D}$ is the translationally invariant dipole
operator, see Eq.~(\ref{dipole_op}) below. The state
$\hat{\bf D}|0_R\rangle$ is a $J^{\pi}=1^-$ state (and similar for the
bra state) and therefore has the same quantum numbers as spurious
center-of-mass excitations.

The coupled-cluster method employs the intrinsic Hamiltonian 
\be
H = T - T_{\rm CoM} + V \ .
\ee
Here, $T$ is the total kinetic energy, $T_{\rm CoM}$ is the kinetic
energy of the center of mass, and $V$ is the translationally invariant
potential. For the intrinsic Hamiltonian, coupled-cluster computations
of ground and excited states avoid center-of-mass admixtures to a good
precision for practical
purposes~\cite{hagen2009a,hagen2010b}. Spurious center-of-mass 
excitations can be identified as described by \textcite{jansen2012}. However, the
coupled-cluster wavefunctions are not simply products of an intrinsic
wavefunctions and a center-of-mass wavefunction. This is problematic
when the Lanczos procedure is applied to the state
$\hat{\bf D}|0_R\rangle$, because any small admixture of
$\hat{\bf D}|0_R\rangle$ with a center-of-mass state gets amplified in
the Lanczos iteration. As a consequence, the diagonalization of the
complex symmetric Lanczos matrix $\bs{M}$ of Eq.~(\ref{m00}) yields a
very low-lying (and spurious) $J^\pi=1^-$ state.  In sufficiently
large model spaces of about 10 oscillator shells or so, this spurious
state is below 1~MeV of excitation energy. The spurious state would be
at exactly zero energy if the factorization of the intrinsic and
center-of-mass wavefunction were perfect in the coupled-cluster
method. 

In order to remove spurious states, we  follow a procedure which is similar to that used to 
remove the elastic contributions in electron scattering~\cite{bacca2007}. 
As it was noticed in~\cite{efros2007}, when using any diagonalization method 
the LIT in Eq.~(\ref{lorenzog}) can be expressed as

\begin{eqnarray}\label{remove} 
L(z)&=&\frac{1}{\pi} {\rm Im}\left\{\sum_{\nu} \frac{ |\bra \varphi_\nu^N | \Theta |0\ket|^2} 
                     {\epsilon_{\nu}^N-z}\right\}\nonumber\\
                    &=& \frac{\Gamma}{\pi}\sum_{\nu} \frac{ |\bra \varphi_\nu^N 
                    | \Theta^\dagger |0\ket|^2} {(\epsilon_{\nu}^N-E_0-\omega_0)^2+\Gamma^2)}\,.               
\end{eqnarray}
Here the $\epsilon_{\nu}^N$ and $\varphi_\nu^N$ are eigenvalues and eigenfunctions of the diagonalized Hamiltonian matrix
(the index $N$ reminds us that both quantities depend on the size of the basis). 
Thus, the LIT is a sum of Lorentzians centered at $\epsilon_{\nu}^N$ and of width $\Gamma$. 
Of course this is also the case when using the Lanczos diagonalization and the similarity transformed Hamiltonian. 
Therefore, a spurious state $\varphi_\nu^N$  can be removed by omitting it in the sum in Eq.~(\ref{remove}).

\section{Validation in $^4$He}
\label{sec:He4}
By reformulating the LIT approach within the CC theory we have
obtained a new method to tackle break-up observables in nuclei.  As we
have already stressed this method is in principle exact and
approximations only enter through truncation of the $T$ operator in
the similarity transformations in Eqs.~(\ref{hbar}) and
(\ref{theta_bar}), and through truncation at a given particle-hole
excitation level in the excited states $\vert \tilde{\Psi}_R\rangle$
and $\langle \tilde{\Psi}_L\vert$ given in Eqs.~(\ref{eom1})
and~(\ref{eomL}). In what follows we will consider an expansion up to
two-particle-two-hole excitations in both the cluster amplitude $T$
and the excitation operators $\mathcal{R}$ and $\mathcal{L}$,
respectively. This approximation is analogous with the standard
equation-of-motion coupled-cluster with singles-and-doubles
excitations (EOM-CCSD) method~\cite{bartlett2007}. In the following we
label our approximation of the LIT-CC equations by
LIT-CCSD. The computational cost of the LIT-CCSD
scheme is the same as that of EOM-CCSD, namely $n_o^2n_u^4$ where
$n_o$ is the number of occupied orbitals and $n_u$ is number of
un-occupied orbitals. In order to reach model-space sizes large enough
to obtain converged results we solve the LIT-CC equations in an
angular momentum coupled scheme~\cite{hagen2008,hagen2010b}. The
EOM-CCSD diagrams and their corresponding angular momentum coupled
algebraic expressions can be found in~\cite{hagen2013c}.

We first want to benchmark this new method with a known solution of
the problem.  For the mass number $A=4$ extensive studies have been
done with the accurate EIHH method~\cite{barnea2001}. By comparing
EIHH and CC results for $^4$He, where the same interaction and
excitation operator are used, we can study the convergence pattern and
assess the accuracy of the approximations introduced in the
LIT-CCSD scheme.

In all the results shown for this benchmark, and in the following
sections, we will use a chiral nucleon-nucleon force derived at
next-to-next-to-next-to-leading order (N$^3$LO)~\cite{entem2003} and
an excitation operator $\hat \Theta$ equal to the third component of
the dipole operator written in a translational invariant form as
in Ref.~\cite{bacca2013} 
\be 
\label{dipole_op}
\hat{\bf D}=\sum_i^A P_i \left({\bf r}_i
  - {\bf R}_{\rm cm} \right) = \sum_i^A \left(P_i - \frac{Z}{A}\right)
{\bf r}_i \,, 
\ee
where $P_i$ projects onto protons. This implies that the excited states
$\vert \tilde{\Psi}_R\rangle$ and $\langle \tilde{\Psi}_L\vert$ in
Eqs.~(\ref{eom1}) and~(\ref{eomL}) carry the quantum numbers $J^\pi,
T_z = 1^-,0$. Furthermore, in the case of non-scalar excitations we
have that $r_0(z)=0 = l_0(z^*)$ in Eqs.~(\ref{eom1}) and~(\ref{eomL}).
\begin{figure}[ht]
\includegraphics[scale=0.4,clip=]{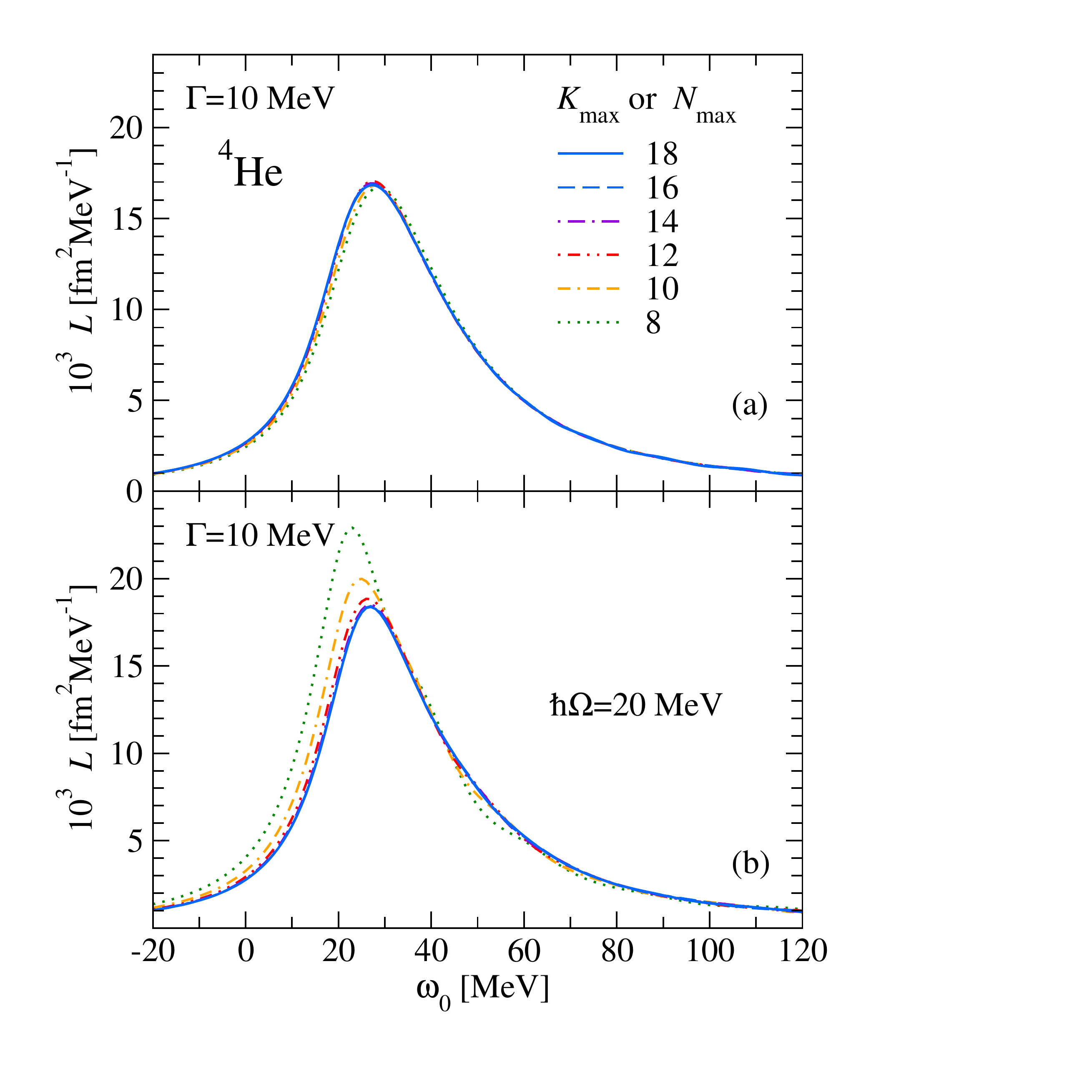} 
\caption{(Color online) Convergence of $L(\omega_0,\Gamma)$ in $^4$He
  with $\Gamma=10$~MeV as a function of $K_{\rm max}$ in the EIHH
  expansion (a).  Convergence of the LIT-CCSD equations with
  $\Gamma=10$~MeV as a function of $N_{\rm max}$ for an HO frequency of
  $\hbar \Omega=20$ MeV (b).}
\label{fig_conv}
\end{figure}

In Figure~\ref{fig_conv}, the LIT of the $^4$He dipole response
function is shown as a function of $\omega_0$ at fixed $\Gamma=10$
MeV. In panel (a) the EIHH results are presented for different model
space sizes, represented by different values of the grandangular
momentum $K_{\rm max}$. The convergence is fast and excellent. In panel
(b) we show the results computed within the LIT-CCSD approach
in model spaces of $N_{\rm max}=2n+l=8, 10, 12, \ldots 18$ and for a
value of the underlying harmonic oscillator (HO) frequency of
$\hbar\Omega = 20$~MeV. Compared to the EIHH calculations, the
LIT-CCSD approach shows a larger difference between the
smallest and largest model space results. However, the LIT is well
converged when $N_{\rm max}=18$ is used and does not change when
varying the underlying HO frequency.
 
\begin{figure}[htb]
  \includegraphics[scale=0.4,clip=]{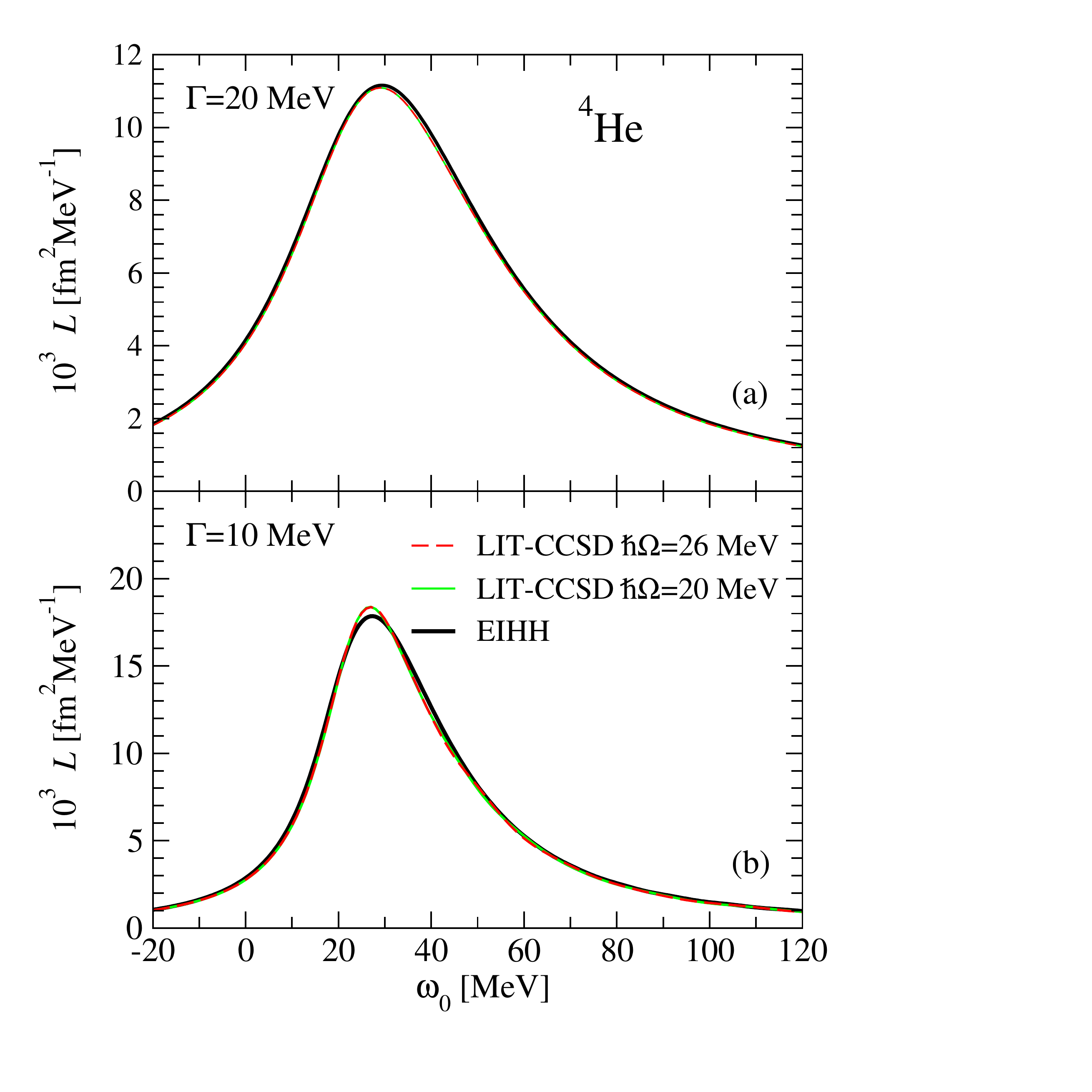}
\caption{(Color online) Comparison of $L(\omega_0,\Gamma)$ in $^4$He
  at $\Gamma=20$ (a), and $\Gamma=10$ (b), calculated in the
  LIT-CCSD scheme with $N_{\rm max}=18$ and two values of
  $\hbar\Omega=20$ and $26$ MeV against the LIT from EIHH.}
\label{fig_comp}
\end{figure}

At this point it is interesting to compare both the EIHH and
LIT-CCSD converged results.  In Figure~\ref{fig_comp}, we
compare the LITs for the values of $\Gamma=20$ and 10~MeV in panel (a)
and (b), respectively. The LIT-CCSD results are shown to
overlap for two values of the harmonic oscillator frequency. They also
agree very well with the EIHH result, especially for $\Gamma=20$
MeV. For the finer resolution scale of $\Gamma=10$~MeV, some slight
differences are observed. It is known that calculations of the LIT
with smaller $\Gamma$ tend to be more cumbersome. In fact as $\Gamma$
decreases the Lorentzian kernel approaches the $\delta-$function,
facing again the continuum problem (for $\omega_0$ above the break-up
threshold). Consequently the Lorentz state approaches the vanishing
boundary condition at further and further distances.  However, since
the convergence of the LIT is very good, as also demonstrated by the
$\hbar \Omega$-independence, we tend to attribute the small
differences with respect to the EIHH result to the truncations
inherent in the LIT-CCSD approximation.
To further quantify the role of the truncation, it is interesting to
compare the dipole response functions obtained by the inversions of
both the calculated LITs.  For the inversions we use the method
outlined in Refs.~\cite{efros1999,andreasi2005}, which looks for the
``regularized solution'' of the integral transform equation.  We
regularize the solution by the following nonlinear ansatz
\be
\label{ansatz}
S(\omega)=\omega^{3/2} \exp \left( -\alpha \pi (Z-1)
  \sqrt{2\mu\over\omega}\right) \sum_i^\nu c_i e^{-\frac{\omega}{i\beta
    }}\,, 
\ee
where $\beta$ is a nonlinear parameters.  Since the first channel
involves the Coulomb force between the emitted proton and the
remaining nucleus with $(Z-1)$ protons a Gamow prefactor is included,
$\alpha$ denotes the fine structure constant, and $\mu$ is the
 reduced mass of the proton and $^{15}N$ system. The coefficients $c_i$ and the parameter $\beta$ are obtained by a
least square fit of the calculated LIT with the integral transform of
the regularized ansatz in Eq.~(\ref{ansatz}), requiring that the
resulting response function is zero below the threshold energy
$\omega_{\rm th}$, where particle emission starts.  For the $^4$He
case, where the first break-up channel is the proton-triton,
$\omega_{\rm th}$ is obtained by the difference of the binding energies of
$^4$He and $^3$H.  The CCSD approximation and the particle-removed
equation-of-motion method \cite{gour2006,hagen2010b} lead to binding
energies $23.97$ and $7.37$~MeV for $^4$He and $^3$H, respectively,
leading to $\omega_{\rm th}=16.60$~MeV.  With the N$^3$LO two-body
interaction precise binding energies are obtained from the EIHH
method and are $25.39$ ($7.85$) MeV for $^4$He ($^3$H), leading to a
slightly different $\omega_{\rm th}=17.54$ MeV.  Because for $^4$He we
know the precise threshold results with the N$^3$LO potential, we
require the response function to be zero below 17.54 MeV, also when we
invert the LIT-CCSD calculations.

\begin{figure}[htb]
  \includegraphics[scale=0.32,clip=]{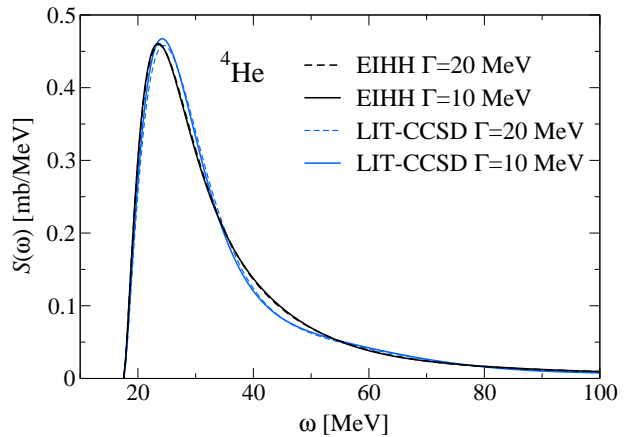}
\caption{(Color online) Comparison of the $^4$He dipole response
  function calculated with LIT-CCSD ($\hbar\Omega=20$ MeV and $N_{\rm
    max}=18$) with the EIHH result. Inversions of the LITs with
  $\Gamma=10$ and 20 are performed.  }
\label{fig_comp_HH_resp}
\end{figure}

 Figure~\ref{fig_comp_HH_resp} shows the comparison of the response
functions obtained by inverting the LIT from the LIT-CCSD and EIHH
methods.  For the LIT-CCSD calculations, we found that the inversions
are insensitive to $\hbar\Omega$.  In principle the inversion should
also not depend on the parameter $\Gamma$. We employ $\Gamma=10$~MeV
and $\Gamma=20$~MeV to gauge the quality of the inversions. For the
EIHH, the inversions obtained from the LITs at $\Gamma=10$ and 20 MeV
overlap very nicely, proving the precision of these calculations.  In
case of the LIT-CCSD, the two values of $\Gamma$ lead to slightly
different inversion, as shown in Figure ~\ref{fig_comp_HH_resp}.  Such
a difference is small and can be viewed as a numerical uncertainty
associated with the inversion.  Overall, the LIT-CCSD response
function is close to the virtually exact EIHH result.  Apparently, the
small deviations between the LIT-CCSD and the EIHH for the LIT in
Figure~\ref{fig_comp} translates into small deviations in the response
function for energies between about $\omega=30$ and $50$~MeV.
\begin{figure}[htb]
  \includegraphics[scale=0.32,clip=]{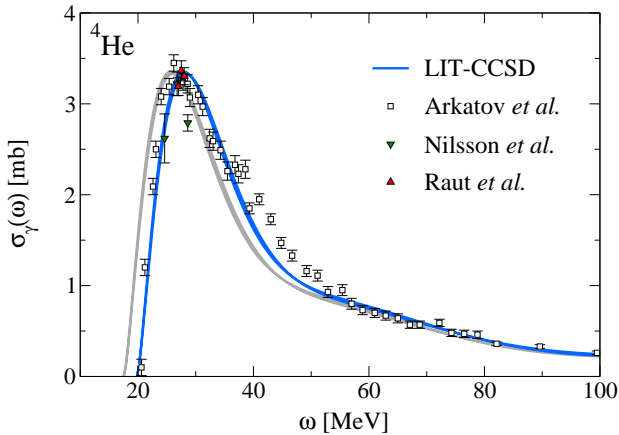}
  \caption{(Color online) Comparison of the $^4$He dipole cross
    section calculated with LIT-CCSD and experimental data from
    Arkatov {\it et al.}~\cite{arkatov}, Nilsson {\it et
      al.}~\cite{nilsson2005} and Raut {\it et al.}~\cite{raut2012}.
    The grey and blue bands differ simply by a shift of the
    theoretical threshold (grey) to the experimental one (dark/blue)
    (see text).}
  \label{fig_comp_exp_He4}
\end{figure}

Finally, for completeness we present a comparison with experimental
data on $^4$He. Extensive studies have been done in the past
concerning the GDR in $^4$He, both from the theoretical and
experimental point of view. Three-nucleon forces are typically
included in the theory (see, {\it e.g.}, Refs.~\cite{gazit2006,
  quaglioni2007}), thus a comparison with data is not conclusive using
two-body forces only. However, even a qualitative comparison is
instructive, especially in the light of addressing heavier nuclei in
the next sessions.

In Figure~\ref{fig_comp_exp_He4} we compare the photoabsorption cross
section calculated in LIT-CCSD (the band width in the theoretical
curves is obtained by filling the difference between the $\Gamma=10$
and 20 MeV inversions) with a selection of the available experimental
data.  The $E1$ photodisintegration cross section is related to the
dipole response as 
\be 
\sigma_{\gamma}^{E1}(\omega)=4\pi^2 \alpha
\omega S(\omega)\,,
\label{cs}
\ee 
with $\alpha$ being the fine structure constant.  Arkatov {\it et
  al.}~\cite{arkatov} measured the photodisintegration cross section
spanning a quite large energy range.  More recent data by Nilsson {\it
  et al.}~\cite{nilsson2005} and Raut {\it et al.}~\cite{raut2012}
cover a narrower range (see Ref.~\cite{bacca2014} for an update on all
the measurements and calculations).  In Figure~\ref{fig_comp_exp_He4},
the grey curve represents the calculation where the theoretical
threshold is used in the inversion.  One notices that this is not as
the experimental one, because the used Hamiltonian misses the
contribution of the three-body force to the binding energies of $^4$He
and$^3$H. Thus, as typically done in the literature, to take this
trivial binding effect into account we shift the theoretical (grey) curve to the experimental threshold (note that the consistent
theoretical threshold is still used in the inversion procedure). It is
evident that the theory describes the experimental data qualitatively,
so it is interesting to address heavier nuclei.

\section{Application to $^{16}$O}
\label{sec:O16} 

The $^4$He benchmark suggests that the LIT-CCSD method can be employed
for the computation of the dipole response, and that theoretical
uncertainties with respect to the model space and the inversion of the
LIT are well controlled. Thus, we turn our attention to a stable
medium-mass nucleus, such as $^{16}$O.

\begin{figure}[htb]
\includegraphics[scale=0.4,clip=]{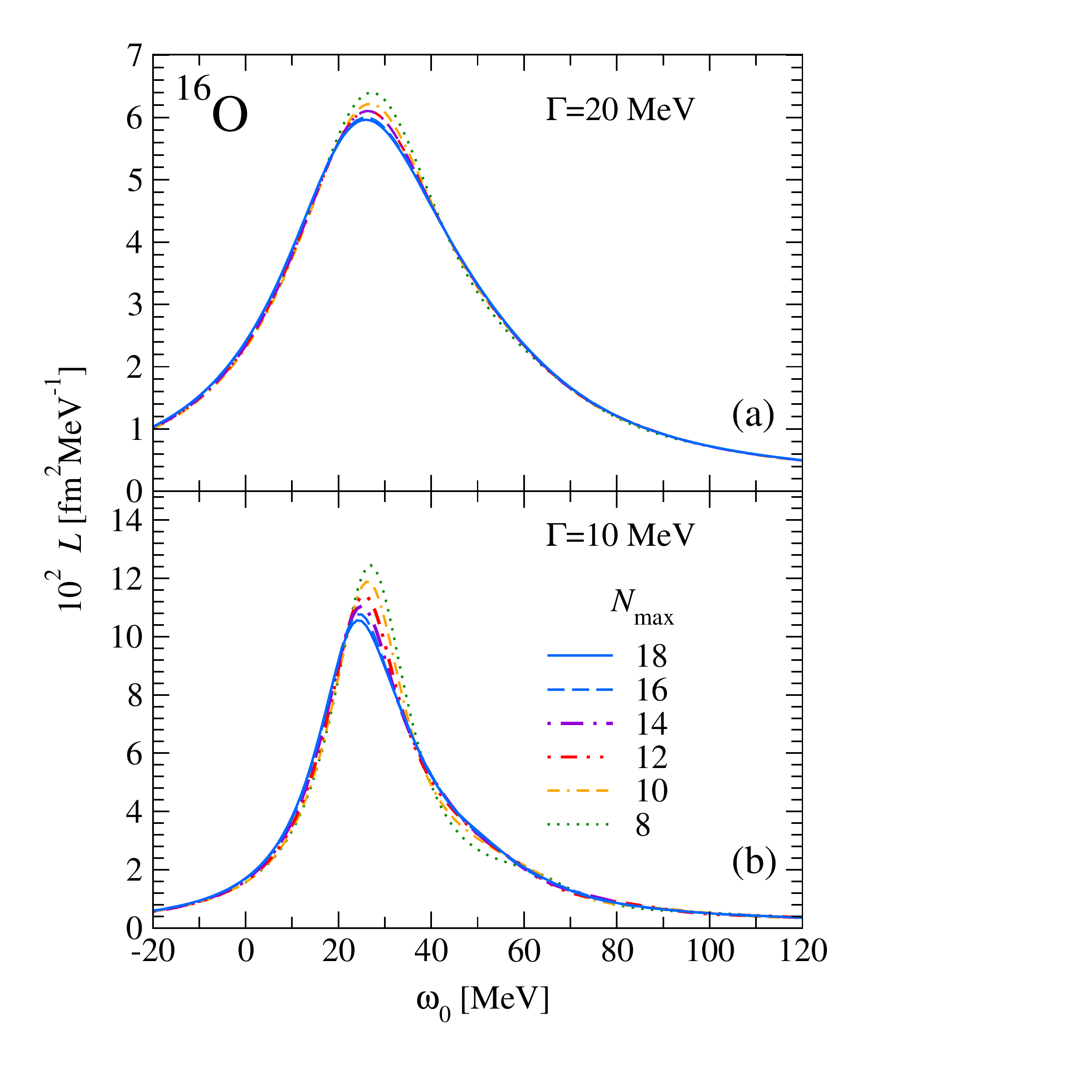}
\caption{(Color online) Convergence of $L(\omega_0,\Gamma)$ in
  $^{16}$O at $\Gamma=20$~MeV (a) and $\Gamma=10$ (b) for different
  values of $N_{\rm max}$ and an HO frequency of $\hbar
  \Omega=26$~MeV.}
\label{fig_conv_O16}
\end{figure}

First, we investigate the convergence of the LIT as a function of the
model space size.  In Figure~\ref{fig_conv_O16}, we present the LITs
for $\Gamma=20$~MeV (panel (a)) and $\Gamma=10$ MeV (panel (b)) with
$N_{\rm max}$ ranging from 8 up to 18.  The convergence is rather good
and it is better for the larger value of $\Gamma$.  As indicated
above, the smaller the width $\Gamma$, the more difficult is to
converge in a LIT calculation.  For $\Gamma=10$ a small difference of
about about $2\%$ between $N_{\rm max}=16$ and $N_{\rm max}=18$ is
found.

\begin{figure}[htb]
  \includegraphics[scale=0.32,clip=]{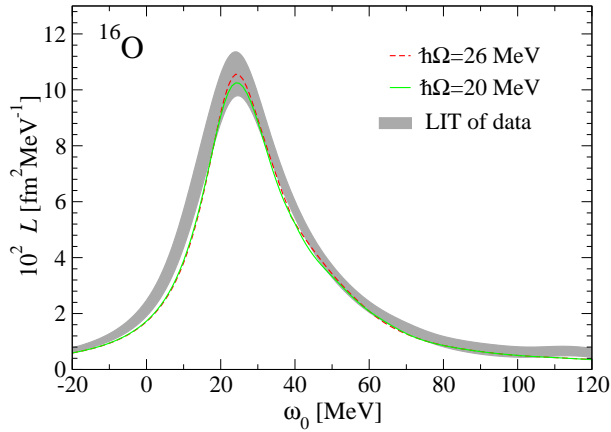}
  \caption{(Color online) Comparison of $L(\omega_0,\Gamma)$ in
    $^{16}$O at $\Gamma=10$ calculated in the LIT-CCSD scheme with
    $N_{\rm max}=18$ and two values of $\hbar\Omega=20$ and $26$ MeV
    against the LIT of the experimental data from Ahrens {\it et
      al.}~\cite{ahrens1975}.}
  \label{fig_litdata_O16}
\end{figure}

Before inverting the transform, it is first interesting to investigate
the $\hbar \Omega$-dependence of our results and compare the theory
with the integral transform of data.  In Figure~\ref{fig_litdata_O16},
LITs from our LIT-CCSD calculations with the largest model space size
of $N_{\rm max}=18$ and two different HO frequencies of
$\hbar\Omega=20$ and $26$~MeV are shown. As one can notice, there is a
residual $\hbar \Omega$ dependence of roughly $4\%$, which is small
and can be considered as the error bar of the numerical calculation.
Overall, the theoretical error associated of our LIT for $\Gamma=10$
MeV in the LIT-CCSD scheme amounts to $5\%$.

The photodisintegration data measured by Ahrens {\it et
  al.}~\cite{ahrens1975} cover a broad energy range. Therefore it is
possible to apply the LIT (Eq.~(\ref{lorenzo})) on the response
function extracted from the data by Eq.~(\ref{cs}).  This allows us to
compare the experimental and theoretical results, as done in
Figure~\ref{fig_litdata_O16} (the area between the grey lines represents the data error
band).  Our theoretical predictions agree with the experimental LIT
within the uncertainties in almost all the $\omega_0$ range.  Only
from $\omega_0=0$ to about $15$~MeV, the theory slightly
underestimates the data.  Since the Lorentzian kernel in
Eq.~(\ref{lorenzo}) is a representation of the $\delta$-function the
integral in $\omega_0$ of $L(\omega_0,\Gamma)$ is the same as the
integral in $\omega$ of $S(\omega)$. Also peak positions are
approximately conserved.  Consequently, from
Figure~\ref{fig_litdata_O16} we can infer that the LIT-CCSD
calculation will reproduce the centroid of the experimental dipole
response and the total strength.

\begin{figure}[htb]
  \includegraphics[scale=0.32,clip=]{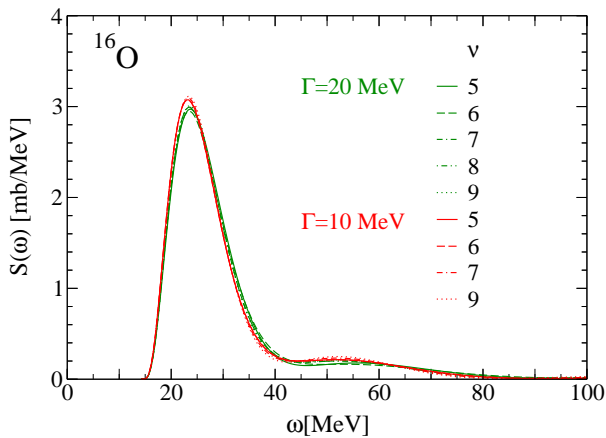}
  \caption{(Color online) The dipole response function of $^{16}$O
    obtained by inverting $L(\omega_0,\Gamma)$ at $\Gamma=20$ and 10
    MeV for different values $\nu$ of the basis functions in
    Eq.~(\ref{ansatz}). }
\label{fig_lit_nudata_O16}
\end{figure}

\begin{figure}[htb]
\includegraphics[scale=0.32,clip=]{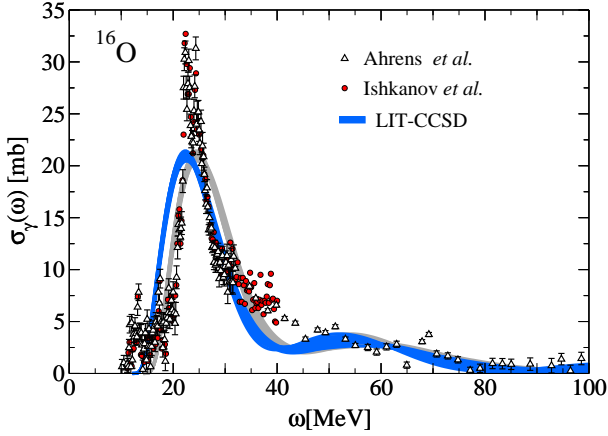}
\caption{(Color online) Comparison of the $^{16}$O dipole cross
  section calculated in the LIT-CCSD scheme against experimental data
  by Ahrens {\it et al.}~\cite{ahrens1975} (triangles with error
  bars), and Ishkhanov {\it et al.}~\cite{ishkhanov2002} (red
  circles).  The grey curve starts from the theoretical threshold,
  while the dark/blue curve is shifted to the experimental threshold.}
\label{fig_resp_O16}
\end{figure}

At this point, we perform the inversion of the computed LIT using the
ansatz of Eq.~(\ref{ansatz}), in order to compare with the cross
section directly. Let us first investigate the stability of the
inversions.  In Figure~\ref{fig_lit_nudata_O16} we show the inversions
of the $^{16}$O LITs at $\Gamma=20$ and 10 MeV for different values $\nu$ of
the basis functions in Eq.~(\ref{ansatz}).  Within the CCSD
scheme, the binding energy of $^{16}$O is 107.24~MeV and with the more
precise perturbative-triples approach, $\Lambda$-CCSD(T)
\cite{taube2008,hagen2010b}, it becomes 121.47~MeV. The threshold
energy, in this case is the difference between the binding energy of
$^{16}$O and $^{15}$N, and is computed using the particle-removed
equation-of-motion theory. For the $^{16}$O photodisintegration
reaction $\omega_{\rm th}$ becomes then $14.25$~MeV and in the
inversion we require the response function to be zero below this
threshold.  For a fixed value of $\Gamma$, several choices of the
number of basis states (from $\nu=5$ to 9) lead to basically the same
inversion. For $^{16}$O the inversions obtained from the LIT at
$\Gamma=10$ MeV are slightly different than those obtained from the
LIT at $\Gamma=20$ MeV. This is due to the fact that the corresponding
LITs themselves are converged only at a few-percent level and not to
the sub-percent level. Because such a difference is very small, we
will interpret it as a numerical error of the inversion and consider a
band made by all of these inversions together as our final result in
the LIT-CCSD scheme. The latter is presented in
Figure~\ref{fig_resp_O16} in comparison to the data by Ahrens {\it et
  al.}~\cite{ahrens1975} and also to the more recent evaluation by
Ishkhanov {\it et al.}~\cite{ishkhanov2002,ishkhanov2004}. The grey
curve represents the LIT-CCSD result plotted starting from the
theoretical threshold and the dark/blue curve is plotted from the
experimental threshold, in analogy to what is done in
Fig.~\ref{fig_comp_exp_He4}.  The position of the GDR in $^{16}$O is
rather well reproduced by our calculations.  We find that the
theoretical width of the GDR is larger than the experimental one,
while the tail region between 40 and 100~MeV is well described within
uncertainties.

In literature, the Thomas-Reiche-Kuhn sum rule is often discussed in relation to the
the photoabsorption cross sections. It is obtained by the following integral
\begin{equation}
\label{TRK}
\int_{\omega_{\rm th}}^{\infty} d\omega \sigma(\omega)=59.74(NZ/A){\rm ~MeV mb~} (1+\kappa) \,
\end{equation}
and $\kappa$ is the so-called enhancement factor. The latter is related to the 
contribution of exchange terms in the nucleon-nucleon force and their induced correlations~\cite{Orlandini91}).
When integrating the theoretical photoabsorption cross section up to
$100$ MeV we obtain an enhancement $\kappa=0.57-0.58$ of the
Thomas-Reiche-Kuhn sum rule $\left[59.74 \frac{NZ}{A}{\rm MeV~
    mb}(1+\kappa)\right]$.

\section{Application to $^{22}$O}
\label{sec:O22}

It is interesting to apply the present method to the study of the
dipole response function of the neutron-rich nucleus $^{22}$O.

\begin{figure}[htb]
\includegraphics[scale=0.333335,clip=]{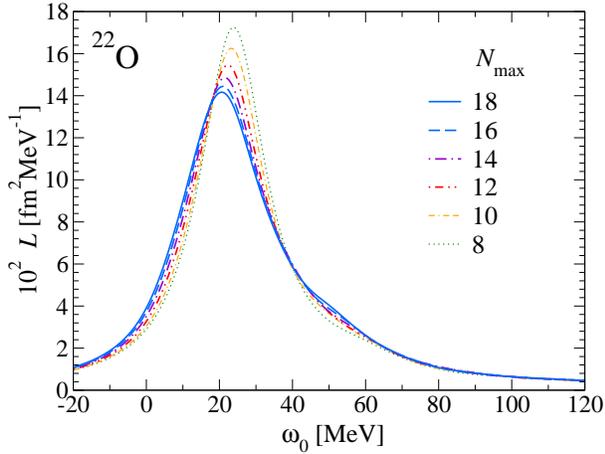}
\caption{(Color online) Convergence of $L(\omega_0,\Gamma)$ in
  $^{22}$O at $\Gamma=10$~MeV as a function of $N_{\rm max}$ for an
  harmonic oscillator frequency of $\hbar \Omega=24$~MeV.}
\label{fig_conv_O22}
\end{figure}

Figure~\ref{fig_conv_O22} shows the convergence of the
LIT, as a function of the model space size, presenting
$L(\omega_0,\Gamma)$ for $\Gamma=10$~MeV with $N_{\rm max}$ ranging
from 8 up to 18.  We observe that the convergence rate is comparable
to that found in $^{16}$O.
\begin{figure}[htb]
\includegraphics[scale=0.333335,clip=]{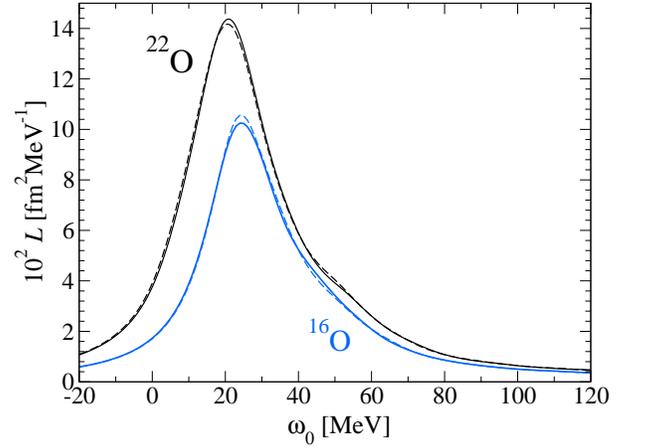}
\caption{(Color online) Comparison of $L(\omega_0,\Gamma)$ at
  $\Gamma=10$ MeV for $^{22}$O and $^{16}$O. Different harmonic
  oscillator frequencies have been used: $\hbar \Omega$ = 20 and 24
  MeV for $^{16}$O (dashed and full blue lines) and $\hbar \Omega$ =
  24 and 26 MeV for $^{22}$O (dashed and full black lines). }
\label{fig_O22_vs_O16}
\end{figure}
\begin{figure}[htb]
\includegraphics[scale=0.333335,clip=]{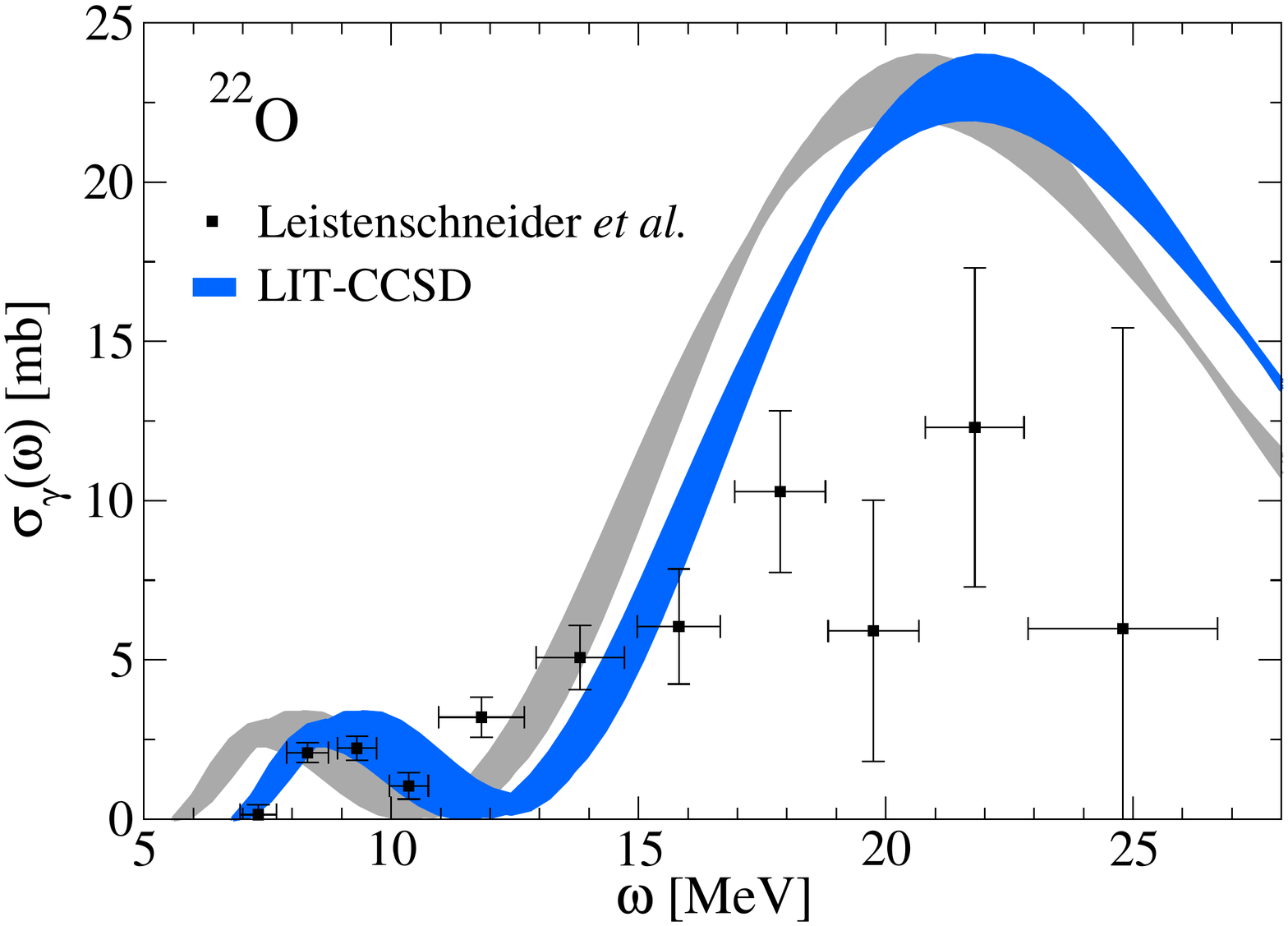}
\caption{(Color online) Comparison of the LIT-CCSD dipole
  cross section of $^{22}$O with the photoneutron data of
  Ref.~\cite{leistenschneider2001}.  The grey curve starts from the
  theoretical threshold, while the dark/blue curve is shifted to the
  experimental threshold.}
\label{22O_comp_data}
\end{figure}

In Fig.~\ref{fig_O22_vs_O16} we compare the LIT for $^{22}$O versus
$^{16}$O for the width $\Gamma=10$ MeV. One notices that the $^{22}$O
total strength is larger than that of $^{16}$O. The total dipole
strength is the bremsstrahlung sum rule (BSR) \be\label{BSR} {\rm BSR}
\equiv \int_{\omega_{\rm th}}^{\infty} d\omega S(\omega)= \langle 0 |
{\hat{D_0}}^{\dagger}\hat{D_0}|0 \rangle \,.  \ee Using the definition
of the LIT, Eq.~(\ref{lorenzo}), and the properties of the Lorentzian
kernel the BSR can also be written as \be\label{BSR2} {\rm BSR} =
\int_{-\infty}^{\infty} d\omega_0 L(\omega_0,\Gamma)\;.  \ee In both
ways we obtain a value of 4.6 and 6.7 fm$^2$ for $^{16}$O and
$^{22}$O, respectively.

We note that the BSR can also be written as~\cite{brink1957}
\be
\label{brink} 
{\rm BSR} \propto
\left(\frac{NZ}{A}\right)^2 {R}_{PN}^2 \ . 
\ee
Here ${R}_{PN}$ is the difference between the proton and the neutron
centers of mass.  If one assumes that the two centers of mass do not
differ much in $^{16}$O and $^{22}$O, difference in the BSR between
$^{16}$O and $^{22}$ O is explained by the different neutron numbers
and mass numbers. This is indeed what we observe within 10\%.

Inverting the LIT and imposing the strength to be zero below the N$^3$LO 
threshold energy of 5.6 MeV, we find the cross section displayed in
Fig.~\ref{22O_comp_data}.  In this case we did not include the Gamow
prefactor of Eq.~(\ref{ansatz}) in the inversion, because the first
channel corresponds to the emission of a neutron.  One notices the
appearance of a small peak at low energy. The existence of such a peak
is a stable feature, independent on the inversion uncertainties.  The
latter are represented by the band width of the curves, obtained by
inverting LITs with $\Gamma=5,10$ and 20 MeV and varying the $\nu$ in
Eq.~(\ref{ansatz}).  As before, the grey curve corresponds to the
LIT-CCSD result starting from the theoretical threshold, while the
dark/blue curve is shifted to the experimental threshold.  After this
shift is performed, it is even more evident that the strength of this
low-lying peak reproduces the experimental one.  Such low-energy peaks
in the dipole response are debated as dipole modes of the excess
neutrons against an $^{16}$O core, see, also Ref.~\cite{repko2013}.
However, like for the experimental result, the strength of this low
energy peak only exhausts about 10\% of the cluster sum
rule~\cite{alhassid1982} inspired by that interpretation.

This is not the first time that the LIT approach suggests the
existence of a low-energy dipole mode. In fact Ref.~\cite{bacca2002}
predicts a similar, but much more pronounced peak in $^6$He for
semirealistic interactions. In that case, however, due to the much
bigger ratio of the neutron halo to the core, the cluster sum rule is
fully exhausted.

When integrating the theoretical photo-absorption cross section up to
$100$ MeV we obtain an enhancement $\kappa=0.54-0.57$ of the
Thomas-Reiche-Kuhn sum rule.

\section{Application to $^{40}$Ca}
\label{sec:Ca40} 
The computational cost of the CC method scales mildly with respect to
the mass number $A$ and the size of the model space. This allows us to
tackle the GDR in $^{40}$Ca, for which data by Ahrens {\it et
  al.}~\cite{ahrens1975} exist from photoabsorption on natural samples
of calcium.
\begin{figure}[htb]
\includegraphics[scale=0.42,clip=]{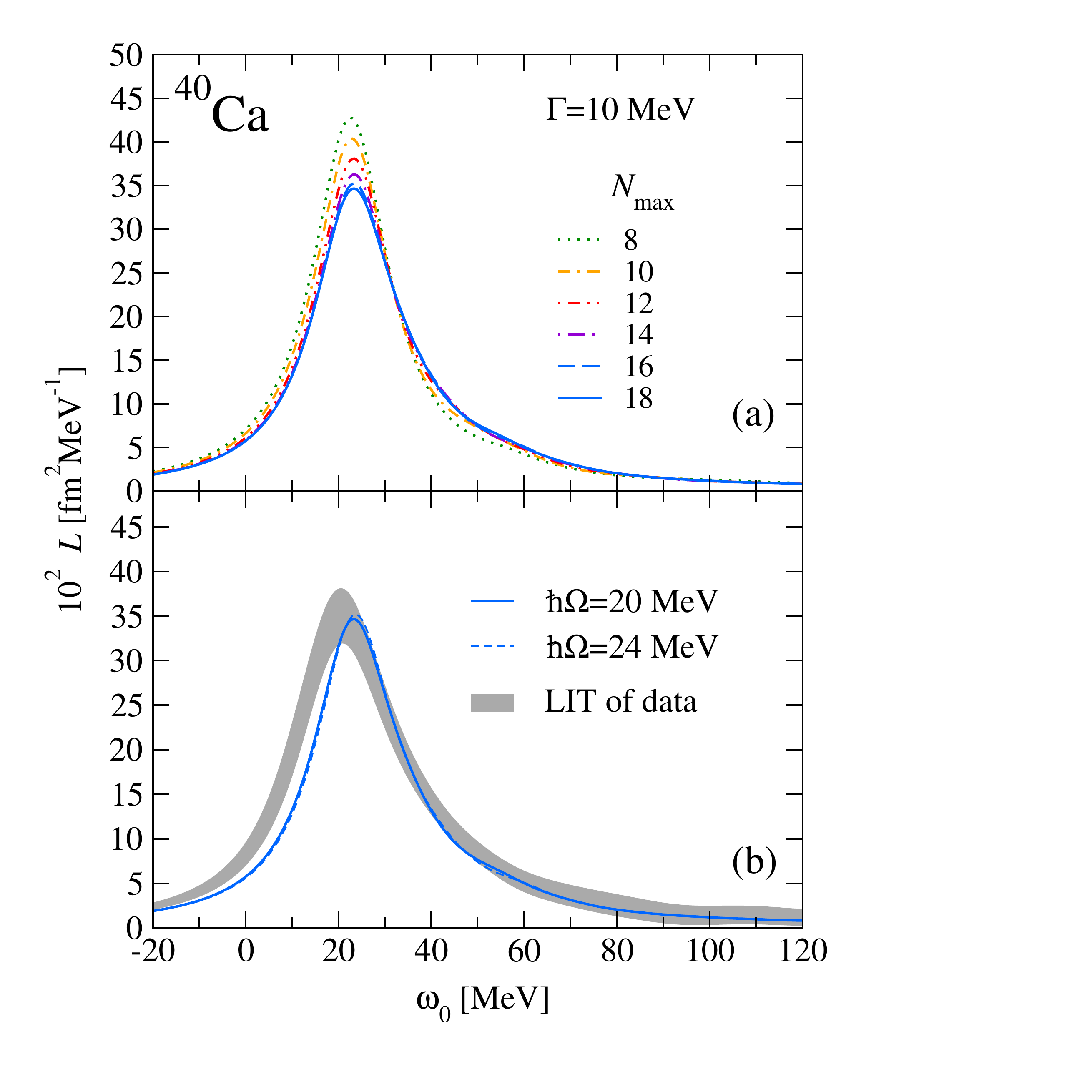}
\caption{(Color online) Convergence of $L(\omega_0,\Gamma)$ in
  $^{40}$Ca at $\Gamma=10$~MeV as a function of $N_{\rm max}$ for an
  harmonic oscillator frequency of $\hbar \Omega=20$~MeV (a).
  Comparison of $L(\omega_0,\Gamma)$ in $^{40}$Ca at $\Gamma=10$
  calculated in the LIT-CCSD scheme with $N_{\rm max}=18$ and
  two values of $\hbar\Omega=20$ and $24$ MeV against the LIT of the
  experimental data from Ahrens {\it et al.}~\cite{ahrens1975} (b).  }
\label{fig_conv_Ca40}
\end{figure}

\begin{figure}[htb]
\includegraphics[scale=0.33,clip=]{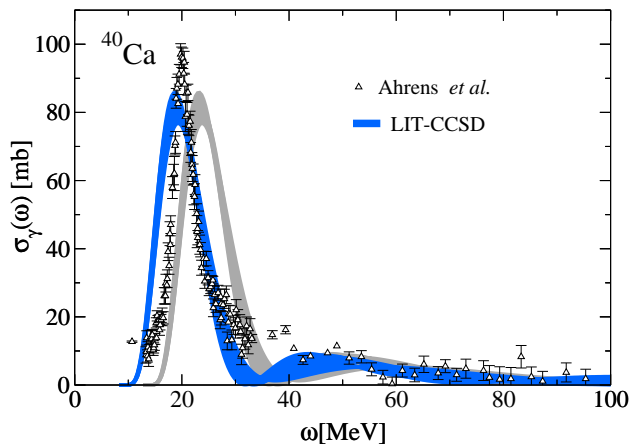}
\caption{(Color online) Comparison of the LIT-CCSD dipole
  cross section of $^{40}$Ca with the photoabsorption data of
  Ref.~\cite{ahrens1975}.  The grey curve starts from the theoretical
  threshold, while the dark/blue curve is shifted to the experimental
  threshold.}
\label{fig_xs_Ca40}
\end{figure}

In Fig.~\ref{fig_conv_Ca40}(a) we show the convergence of the LIT
calculations as a function of $N_{\rm max}$ for a fixed value of the
HO frequency $\hbar \Omega=20$ MeV and for $\Gamma=10$ MeV. It is
apparent that the convergence is of the same quality as for the oxygen
isotopes.  In the bottom panel, a comparison of two LITs with
different underlying HO parameter ($\hbar \Omega=20$ and 24 MeV) is
presented, indicating that the residual $\hbar \Omega$-dependence is
small.  A comparison with the LIT of the experimental data by Ahrens
{\it et al.} is also shown, where the error is represented by the
bands. For $^{40}$Ca the location of the GDR predicted using the N$^3$LO
nucleon-nucleon interaction is found at slightly larger excitation
energy with respect to the experiment. This feature is also reflected
when the LIT is inverted and the photoabsorption cross section is
calculated in the dipole approximation, using Eq.~(\ref{cs}). In this
calculation we apply the ansatz of Eq.~(\ref{ansatz}) using the
threshold energy $\omega_{\rm th}=12.8$ MeV obtained with the
particle-removed equation-of-motion method \cite{gour2006,hagen2010b}.
By taking different widths of the LIT to invert ($\Gamma=5, 10$ and 20
MeV) and by varying the number $\nu$ in Eq.~(\ref{ansatz}), we obtain
the grey band in Fig.~\ref{fig_xs_Ca40}. In comparison to the cross
section data by Ahrens {\it et al.}~\cite{ahrens1975}, the theoretical
prediction of the GDR is quite encouraging. A giant resonance is
clearly seen. However, it is slightly broader, lower in strength and
situated at higher energy than the experimental GDR.  Because the
threshold energy with the N$^3$LO nucleon-nucleon interaction is quite
different than the experimental threshold, located at about 8.3 MeV,
in Fig.~\ref{fig_xs_Ca40}, we also show (in dark blue) the theoretical
curves shifted on the experimental threshold energy.
When integrating the theoretical photo-absorption cross section up to
$100$ MeV we obtain an enhancement $\kappa=0.69-0.73$ of the
Thomas-Reiche-Kuhn sum rule.

Let us also consider the dipole polarizability because of its
considerable experimental and theoretical
interest~\cite{tamii2011,piekarewicz2012}. From the dipole response
function $S(\omega)$ one can obtain the electric dipole polarizability
\begin{equation}
\alpha_E=2\alpha \int_{\omega_{\rm th}}^{\infty} d\omega \frac{S(\omega)}{\omega}
\end{equation}
as an inverse energy weighted sum rule.  In analogy to
Ref.~\cite{tetraedro}, electric dipole polarizability can be also
obtained directly from the Lanczos  approach 
~\cite{goerke2012,Mirko,LanczosSumRules}, avoiding the
inversion of the integral transform. The removal of  center of mass spuriosities for this observable 
can be done in the same way as explained in Section~\ref{sec:remcm}. 
In this case 
\begin{equation}
\alpha_E=2\alpha  \sum_{\nu} \frac{ |\bra \varphi_\nu^N | \Theta |0\ket|^2} 
                     {\epsilon_{\nu}^N}
\end{equation}
and the spurious states can be removed from the sum.
Both from the Lanczos approach
and integrating the response function up to 100 MeV we
obtain $\alpha_E=1.47$~fm$^3$ within $5\%$.  With the present N$^3$LO
nucleon-nucleon interaction we predict a polarizability for $^{40}$Ca,
which is rather low in comparison to the experimental value of
$\alpha^{\rm exp}_E=2.23(3)$ fm$^3$~\cite{ahrens1975}.  If we
integrate the strength after shifting it to the experimental threshold
(dark/blue curve in Fig.~\ref{fig_xs_Ca40}) we obtain roughly
$\alpha_E=1.82$~fm$^3$, thus moving in the direction of the
experimental value. We also note that if we integrate the cross
section data by Ahrens {\it et al.}~\cite{ahrens1975} we obtain
$1.95(26)$~fm$^3$ for the dipole polarizability.  It is worth to
mention that with the present nucleon-nucleon interaction $^{40}$Ca is
about $20$~MeV overbound and with a charge radius $R_{\rm
  ch}=3.05$~fm, which is considerably smaller than the experimental
value of $3.4776(19)$~fm~\cite{angeli2013}. This points towards a
general problem of the present Hamiltonian, which does not provide
good saturation properties of nuclei, leading to too small radii and
consequently too small polarizabilities.

\section{Conclusions}
\label{sec:conclusions}

We presented in detail an approach that combines the Lorentz integral
transform with the coupled-cluster method, named LIT-CC, for the
computation of the dipole response function in $^4$He, $^{16,22}$O and
$^{40}$Ca. The benchmark of this method against the EIHH in $^4$He
gives us the necessary confidence for the computation in heavier
nuclei. The LIT-CCSD approximation yielded results for the total
photonuclear dipole cross section of oxygen and calcium isotopes that
are in semi-quantitative agreement with data. This opens the way for
interesting investigations of the response functions of heavier
nuclei, also beyond the stability valley. 

The comparison of the LITs of the response functions of $^{16}$O and
$^{22}$O shows a larger total area of the latter (corresponding to the
relative bremsstrahlung sum rule) and a slight shift of the peak to
lower energy. Such a shift already envisages the possibility of more
strength in that region.  This becomes manifest after the inversion.
For $^{22}$O we found a very interesting dipole cross section 
exhibiting two peaks:
A small one at 8-9 MeV and a larger one at 21-22 MeV. We also extend our
calculations further out in mass number, presenting first results on
the GDR of $^{40}$Ca.  In this case we observe that, with respect to
experiment, the N$^3$LO nucleon-nucleon interaction leads to larger
excitation energy of the GDR, which is consistent with the
over-binding, the too small charge radius and dipole polarizability we
obtain for $^{40}$Ca.  The results presented here also open the way to
systematic investigations of more general electro-weak responses of
medium-mass nuclei with an {\it ab-initio} approach.

\begin{acknowledgments}
  This work was supported in parts by the Natural Sciences and
  Engineering Research Council (NSERC), the National Research Council
  of Canada, the US-Israel Binational Science Foundation (Grant
  No.~2012212), the Pazy Foundation, the MIUR grant PRIN-2009TWL3MX,
  the Office of Nuclear Physics, U.S.~Department of Energy under
  Grants Nos.~DE-FG02-96ER40963 (University of Tennessee) and
  DE-SC0008499 (NUCLEI SciDAC collaboration), and the Field Work
  Proposal ERKBP57 at Oak Ridge National Laboratory. Computer time
  was provided by the Innovative and Novel Computational Impact on
  Theory and Experiment (INCITE) program. This research used resources
  of the Oak Ridge Leadership Computing Facility located in the Oak
  Ridge National Laboratory, supported by the Office of Science of the
  U.S.~Department of Energy under Contract No.  DE-AC05-00OR22725, and
  computational resources of the National Center for Computational
  Sciences, the National Institute for Computational Sciences, and
  TRIUMF.
\end{acknowledgments}

\bibliography{refs_G,refs}

\begin{thebibliography}{77}%
\makeatletter
\providecommand \@ifxundefined [1]{%
 \@ifx{#1\undefined}
}%
\providecommand \@ifnum [1]{%
 \ifnum #1\expandafter \@firstoftwo
 \else \expandafter \@secondoftwo
 \fi
}%
\providecommand \@ifx [1]{%
 \ifx #1\expandafter \@firstoftwo
 \else \expandafter \@secondoftwo
 \fi
}%
\providecommand \natexlab [1]{#1}%
\providecommand \enquote  [1]{``#1''}%
\providecommand \bibnamefont  [1]{#1}%
\providecommand \bibfnamefont [1]{#1}%
\providecommand \citenamefont [1]{#1}%
\providecommand \href@noop [0]{\@secondoftwo}%
\providecommand \href [0]{\begingroup \@sanitize@url \@href}%
\providecommand \@href[1]{\@@startlink{#1}\@@href}%
\providecommand \@@href[1]{\endgroup#1\@@endlink}%
\providecommand \@sanitize@url [0]{\catcode `\\12\catcode `\$12\catcode
  `\&12\catcode `\#12\catcode `\^12\catcode `\_12\catcode `\%12\relax}%
\providecommand \@@startlink[1]{}%
\providecommand \@@endlink[0]{}%
\providecommand \url  [0]{\begingroup\@sanitize@url \@url }%
\providecommand \@url [1]{\endgroup\@href {#1}{\urlprefix }}%
\providecommand \urlprefix  [0]{URL }%
\providecommand \Eprint [0]{\href }%
\providecommand \doibase [0]{http://dx.doi.org/}%
\providecommand \selectlanguage [0]{\@gobble}%
\providecommand \bibinfo  [0]{\@secondoftwo}%
\providecommand \bibfield  [0]{\@secondoftwo}%
\providecommand \translation [1]{[#1]}%
\providecommand \BibitemOpen [0]{}%
\providecommand \bibitemStop [0]{}%
\providecommand \bibitemNoStop [0]{.\EOS\space}%
\providecommand \EOS [0]{\spacefactor3000\relax}%
\providecommand \BibitemShut  [1]{\csname bibitem#1\endcsname}%
\let\auto@bib@innerbib\@empty
\bibitem [{\citenamefont {Baldwin}\ and\ \citenamefont
  {Klaiber}(1947)}]{BaK47}%
  \BibitemOpen
  \bibfield  {author} {\bibinfo {author} {\bibfnamefont {G.~C.}\ \bibnamefont
  {Baldwin}}\ and\ \bibinfo {author} {\bibfnamefont {G.~S.}\ \bibnamefont
  {Klaiber}},\ }\href {\doibase 10.1103/PhysRev.71.3} {\bibfield  {journal}
  {\bibinfo  {journal} {Phys. Rev.}\ }\textbf {\bibinfo {volume} {71}},\
  \bibinfo {pages} {3} (\bibinfo {year} {1947})}\BibitemShut {NoStop}%
\bibitem [{\citenamefont {Leistenschneider}\ \emph {et~al.}(2001)\citenamefont
  {Leistenschneider}, \citenamefont {Aumann}, \citenamefont {Boretzky},
  \citenamefont {Cortina}, \citenamefont {Cub}, \citenamefont {Pramanik},
  \citenamefont {Dostal}, \citenamefont {Elze}, \citenamefont {Emling},
  \citenamefont {Geissel}, \citenamefont {Gr\"unschlo\ss{}}, \citenamefont
  {Hellstr}, \citenamefont {Holzmann}, \citenamefont {Ilievski}, \citenamefont
  {Iwasa}, \citenamefont {Kaspar}, \citenamefont {Kleinb\"ohl}, \citenamefont
  {Kratz}, \citenamefont {Kulessa}, \citenamefont {Leifels}, \citenamefont
  {Lubkiewicz}, \citenamefont {M\"unzenberg}, \citenamefont {Reiter},
  \citenamefont {Rejmund}, \citenamefont {Scheidenberger}, \citenamefont
  {Schlegel}, \citenamefont {Simon}, \citenamefont {Stroth}, \citenamefont
  {S\"ummerer}, \citenamefont {Wajda}, \citenamefont {Wal\'us},\ and\
  \citenamefont {Wan}}]{leistenschneider2001}%
  \BibitemOpen
  \bibfield  {author} {\bibinfo {author} {\bibfnamefont {A.}~\bibnamefont
  {Leistenschneider}}, \bibinfo {author} {\bibfnamefont {T.}~\bibnamefont
  {Aumann}}, \bibinfo {author} {\bibfnamefont {K.}~\bibnamefont {Boretzky}},
  \bibinfo {author} {\bibfnamefont {D.}~\bibnamefont {Cortina}}, \bibinfo
  {author} {\bibfnamefont {J.}~\bibnamefont {Cub}}, \bibinfo {author}
  {\bibfnamefont {U.~D.}\ \bibnamefont {Pramanik}}, \bibinfo {author}
  {\bibfnamefont {W.}~\bibnamefont {Dostal}}, \bibinfo {author} {\bibfnamefont
  {T.~W.}\ \bibnamefont {Elze}}, \bibinfo {author} {\bibfnamefont
  {H.}~\bibnamefont {Emling}}, \bibinfo {author} {\bibfnamefont
  {H.}~\bibnamefont {Geissel}}, \bibinfo {author} {\bibfnamefont
  {A.}~\bibnamefont {Gr\"unschlo\ss{}}}, \bibinfo {author} {\bibfnamefont
  {M.}~\bibnamefont {Hellstr}}, \bibinfo {author} {\bibfnamefont
  {R.}~\bibnamefont {Holzmann}}, \bibinfo {author} {\bibfnamefont
  {S.}~\bibnamefont {Ilievski}}, \bibinfo {author} {\bibfnamefont
  {N.}~\bibnamefont {Iwasa}}, \bibinfo {author} {\bibfnamefont
  {M.}~\bibnamefont {Kaspar}}, \bibinfo {author} {\bibfnamefont
  {A.}~\bibnamefont {Kleinb\"ohl}}, \bibinfo {author} {\bibfnamefont {J.~V.}\
  \bibnamefont {Kratz}}, \bibinfo {author} {\bibfnamefont {R.}~\bibnamefont
  {Kulessa}}, \bibinfo {author} {\bibfnamefont {Y.}~\bibnamefont {Leifels}},
  \bibinfo {author} {\bibfnamefont {E.}~\bibnamefont {Lubkiewicz}}, \bibinfo
  {author} {\bibfnamefont {G.}~\bibnamefont {M\"unzenberg}}, \bibinfo {author}
  {\bibfnamefont {P.}~\bibnamefont {Reiter}}, \bibinfo {author} {\bibfnamefont
  {M.}~\bibnamefont {Rejmund}}, \bibinfo {author} {\bibfnamefont
  {C.}~\bibnamefont {Scheidenberger}}, \bibinfo {author} {\bibfnamefont
  {C.}~\bibnamefont {Schlegel}}, \bibinfo {author} {\bibfnamefont
  {H.}~\bibnamefont {Simon}}, \bibinfo {author} {\bibfnamefont
  {J.}~\bibnamefont {Stroth}}, \bibinfo {author} {\bibfnamefont
  {K.}~\bibnamefont {S\"ummerer}}, \bibinfo {author} {\bibfnamefont
  {E.}~\bibnamefont {Wajda}}, \bibinfo {author} {\bibfnamefont
  {W.}~\bibnamefont {Wal\'us}}, \ and\ \bibinfo {author} {\bibfnamefont
  {S.}~\bibnamefont {Wan}},\ }\href {\doibase 10.1103/PhysRevLett.86.5442}
  {\bibfield  {journal} {\bibinfo  {journal} {Phys. Rev. Lett.}\ }\textbf
  {\bibinfo {volume} {86}},\ \bibinfo {pages} {5442} (\bibinfo {year}
  {2001})}\BibitemShut {NoStop}%
\bibitem [{\citenamefont {K{\"u}mmel}\ \emph {et~al.}(1978)\citenamefont
  {K{\"u}mmel}, \citenamefont {L{\"u}hrmann},\ and\ \citenamefont
  {Zabolitzky}}]{kuemmel1978}%
  \BibitemOpen
  \bibfield  {author} {\bibinfo {author} {\bibfnamefont {H.}~\bibnamefont
  {K{\"u}mmel}}, \bibinfo {author} {\bibfnamefont {K.~H.}\ \bibnamefont
  {L{\"u}hrmann}}, \ and\ \bibinfo {author} {\bibfnamefont {J.~G.}\
  \bibnamefont {Zabolitzky}},\ }\href {\doibase 10.1016/0370-1573(78)90081-9}
  {\bibfield  {journal} {\bibinfo  {journal} {Physics Reports}\ }\textbf
  {\bibinfo {volume} {36}},\ \bibinfo {pages} {1 } (\bibinfo {year}
  {1978})}\BibitemShut {NoStop}%
\bibitem [{\citenamefont {Bishop}(1991)}]{bishop1991}%
  \BibitemOpen
  \bibfield  {author} {\bibinfo {author} {\bibfnamefont {R.~F.}\ \bibnamefont
  {Bishop}},\ }\href {http://dx.doi.org/10.1007/BF01119617} {\bibfield
  {journal} {\bibinfo  {journal} {Theoretical Chemistry Accounts: Theory,
  Computation, and Modeling (Theoretica Chimica Acta)}\ }\textbf {\bibinfo
  {volume} {80}},\ \bibinfo {pages} {95} (\bibinfo {year} {1991})},\ \bibinfo
  {note} {10.1007/BF01119617}\BibitemShut {NoStop}%
\bibitem [{\citenamefont {Hagen}\ \emph {et~al.}(2013)\citenamefont {Hagen},
  \citenamefont {Hagen}, \citenamefont {Hammer},\ and\ \citenamefont
  {Platter}}]{hagen2013}%
  \BibitemOpen
  \bibfield  {author} {\bibinfo {author} {\bibfnamefont {G.}~\bibnamefont
  {Hagen}}, \bibinfo {author} {\bibfnamefont {P.}~\bibnamefont {Hagen}},
  \bibinfo {author} {\bibfnamefont {H.-W.}\ \bibnamefont {Hammer}}, \ and\
  \bibinfo {author} {\bibfnamefont {L.}~\bibnamefont {Platter}},\ }\href
  {\doibase 10.1103/PhysRevLett.111.132501} {\bibfield  {journal} {\bibinfo
  {journal} {Phys. Rev. Lett.}\ }\textbf {\bibinfo {volume} {111}},\ \bibinfo
  {pages} {132501} (\bibinfo {year} {2013})}\BibitemShut {NoStop}%
\bibitem [{\citenamefont {Tsukiyama}\ \emph {et~al.}(2011)\citenamefont
  {Tsukiyama}, \citenamefont {Bogner},\ and\ \citenamefont
  {Schwenk}}]{tsukiyama2011}%
  \BibitemOpen
  \bibfield  {author} {\bibinfo {author} {\bibfnamefont {K.}~\bibnamefont
  {Tsukiyama}}, \bibinfo {author} {\bibfnamefont {S.~K.}\ \bibnamefont
  {Bogner}}, \ and\ \bibinfo {author} {\bibfnamefont {A.}~\bibnamefont
  {Schwenk}},\ }\href {\doibase 10.1103/PhysRevLett.106.222502} {\bibfield
  {journal} {\bibinfo  {journal} {Phys. Rev. Lett.}\ }\textbf {\bibinfo
  {volume} {106}},\ \bibinfo {pages} {222502} (\bibinfo {year}
  {2011})}\BibitemShut {NoStop}%
\bibitem [{\citenamefont {Hergert}\ \emph {et~al.}(2013)\citenamefont
  {Hergert}, \citenamefont {Binder}, \citenamefont {Calci}, \citenamefont
  {Langhammer},\ and\ \citenamefont {Roth}}]{hergert2013}%
  \BibitemOpen
  \bibfield  {author} {\bibinfo {author} {\bibfnamefont {H.}~\bibnamefont
  {Hergert}}, \bibinfo {author} {\bibfnamefont {S.}~\bibnamefont {Binder}},
  \bibinfo {author} {\bibfnamefont {A.}~\bibnamefont {Calci}}, \bibinfo
  {author} {\bibfnamefont {J.}~\bibnamefont {Langhammer}}, \ and\ \bibinfo
  {author} {\bibfnamefont {R.}~\bibnamefont {Roth}},\ }\href {\doibase
  10.1103/PhysRevLett.110.242501} {\bibfield  {journal} {\bibinfo  {journal}
  {Phys. Rev. Lett.}\ }\textbf {\bibinfo {volume} {110}},\ \bibinfo {pages}
  {242501} (\bibinfo {year} {2013})}\BibitemShut {NoStop}%
\bibitem [{\citenamefont {Dickhoff}\ and\ \citenamefont
  {Barbieri}(2004)}]{dickhoff2004}%
  \BibitemOpen
  \bibfield  {author} {\bibinfo {author} {\bibfnamefont {W.}~\bibnamefont
  {Dickhoff}}\ and\ \bibinfo {author} {\bibfnamefont {C.}~\bibnamefont
  {Barbieri}},\ }\href {\doibase 10.1016/j.ppnp.2004.02.038} {\bibfield
  {journal} {\bibinfo  {journal} {Progress in Particle and Nuclear Physics}\
  }\textbf {\bibinfo {volume} {52}},\ \bibinfo {pages} {377 } (\bibinfo {year}
  {2004})}\BibitemShut {NoStop}%
\bibitem [{\citenamefont {Som\`a}\ \emph {et~al.}(2013)\citenamefont {Som\`a},
  \citenamefont {Barbieri},\ and\ \citenamefont {Duguet}}]{soma2013}%
  \BibitemOpen
  \bibfield  {author} {\bibinfo {author} {\bibfnamefont {V.}~\bibnamefont
  {Som\`a}}, \bibinfo {author} {\bibfnamefont {C.}~\bibnamefont {Barbieri}}, \
  and\ \bibinfo {author} {\bibfnamefont {T.}~\bibnamefont {Duguet}},\ }\href
  {\doibase 10.1103/PhysRevC.87.011303} {\bibfield  {journal} {\bibinfo
  {journal} {Phys. Rev. C}\ }\textbf {\bibinfo {volume} {87}},\ \bibinfo
  {pages} {011303} (\bibinfo {year} {2013})}\BibitemShut {NoStop}%
\bibitem [{\citenamefont {L{\"a}hde}\ \emph {et~al.}(2014)\citenamefont
  {L{\"a}hde}, \citenamefont {Epelbaum}, \citenamefont {Krebs}, \citenamefont
  {Lee}, \citenamefont {Mei{\ss}ner},\ and\ \citenamefont
  {Rupak}}]{laehde2014}%
  \BibitemOpen
  \bibfield  {author} {\bibinfo {author} {\bibfnamefont {T.~A.}\ \bibnamefont
  {L{\"a}hde}}, \bibinfo {author} {\bibfnamefont {E.}~\bibnamefont {Epelbaum}},
  \bibinfo {author} {\bibfnamefont {H.}~\bibnamefont {Krebs}}, \bibinfo
  {author} {\bibfnamefont {D.}~\bibnamefont {Lee}}, \bibinfo {author}
  {\bibfnamefont {U.-G.}\ \bibnamefont {Mei{\ss}ner}}, \ and\ \bibinfo {author}
  {\bibfnamefont {G.}~\bibnamefont {Rupak}},\ }\href {\doibase
  10.1016/j.physletb.2014.03.023} {\bibfield  {journal} {\bibinfo  {journal}
  {Physics Letters B}\ }\textbf {\bibinfo {volume} {732}},\ \bibinfo {pages}
  {110 } (\bibinfo {year} {2014})}\BibitemShut {NoStop}%
\bibitem [{\citenamefont {Hagen}\ \emph {et~al.}(2007)\citenamefont {Hagen},
  \citenamefont {Dean}, \citenamefont {Hjorth-Jensen},\ and\ \citenamefont
  {Papenbrock}}]{hagen2007d}%
  \BibitemOpen
  \bibfield  {author} {\bibinfo {author} {\bibfnamefont {G.}~\bibnamefont
  {Hagen}}, \bibinfo {author} {\bibfnamefont {D.~J.}\ \bibnamefont {Dean}},
  \bibinfo {author} {\bibfnamefont {M.}~\bibnamefont {Hjorth-Jensen}}, \ and\
  \bibinfo {author} {\bibfnamefont {T.}~\bibnamefont {Papenbrock}},\ }\href
  {\doibase 10.1016/j.physletb.2007.07.072} {\bibfield  {journal} {\bibinfo
  {journal} {Physics Letters B}\ }\textbf {\bibinfo {volume} {656}},\ \bibinfo
  {pages} {169 } (\bibinfo {year} {2007})}\BibitemShut {NoStop}%
\bibitem [{\citenamefont {Hagen}\ \emph
  {et~al.}(2012{\natexlab{a}})\citenamefont {Hagen}, \citenamefont
  {Hjorth-Jensen}, \citenamefont {Jansen}, \citenamefont {Machleidt},\ and\
  \citenamefont {Papenbrock}}]{hagen2012b}%
  \BibitemOpen
  \bibfield  {author} {\bibinfo {author} {\bibfnamefont {G.}~\bibnamefont
  {Hagen}}, \bibinfo {author} {\bibfnamefont {M.}~\bibnamefont
  {Hjorth-Jensen}}, \bibinfo {author} {\bibfnamefont {G.~R.}\ \bibnamefont
  {Jansen}}, \bibinfo {author} {\bibfnamefont {R.}~\bibnamefont {Machleidt}}, \
  and\ \bibinfo {author} {\bibfnamefont {T.}~\bibnamefont {Papenbrock}},\
  }\href {\doibase 10.1103/PhysRevLett.109.032502} {\bibfield  {journal}
  {\bibinfo  {journal} {Phys. Rev. Lett.}\ }\textbf {\bibinfo {volume} {109}},\
  \bibinfo {pages} {032502} (\bibinfo {year} {2012}{\natexlab{a}})}\BibitemShut
  {NoStop}%
\bibitem [{\citenamefont {Hagen}\ and\ \citenamefont
  {Michel}(2012)}]{hagen2012c}%
  \BibitemOpen
  \bibfield  {author} {\bibinfo {author} {\bibfnamefont {G.}~\bibnamefont
  {Hagen}}\ and\ \bibinfo {author} {\bibfnamefont {N.}~\bibnamefont {Michel}},\
  }\href {\doibase 10.1103/PhysRevC.86.021602} {\bibfield  {journal} {\bibinfo
  {journal} {Phys. Rev. C}\ }\textbf {\bibinfo {volume} {86}},\ \bibinfo
  {pages} {021602} (\bibinfo {year} {2012})}\BibitemShut {NoStop}%
\bibitem [{\citenamefont {Deltuva}\ and\ \citenamefont
  {Fonseca}(2012)}]{DeF12}%
  \BibitemOpen
  \bibfield  {author} {\bibinfo {author} {\bibfnamefont {A.}~\bibnamefont
  {Deltuva}}\ and\ \bibinfo {author} {\bibfnamefont {A.~C.}\ \bibnamefont
  {Fonseca}},\ }\href@noop {} {\bibfield  {journal} {\bibinfo  {journal} {Phys.
  Rev. C}\ }\textbf {\bibinfo {volume} {86}},\ \bibinfo {pages} {011001}
  (\bibinfo {year} {2012})}\BibitemShut {NoStop}%
\bibitem [{\citenamefont {Efros}\ \emph {et~al.}(1994)\citenamefont {Efros},
  \citenamefont {Leidemann},\ and\ \citenamefont {Orlandini}}]{efros1994}%
  \BibitemOpen
  \bibfield  {author} {\bibinfo {author} {\bibfnamefont {V.~D.}\ \bibnamefont
  {Efros}}, \bibinfo {author} {\bibfnamefont {W.}~\bibnamefont {Leidemann}}, \
  and\ \bibinfo {author} {\bibfnamefont {G.}~\bibnamefont {Orlandini}},\ }\href
  {\doibase 10.1016/0370-2693(94)91355-2} {\bibfield  {journal} {\bibinfo
  {journal} {Physics Letters B}\ }\textbf {\bibinfo {volume} {338}},\ \bibinfo
  {pages} {130 } (\bibinfo {year} {1994})}\BibitemShut {NoStop}%
\bibitem [{\citenamefont {Martinelli}\ \emph {et~al.}(1995)\citenamefont
  {Martinelli}, \citenamefont {Kamada}, \citenamefont {Orlandini},\ and\
  \citenamefont {Gl\"ockle}}]{martinelli1995}%
  \BibitemOpen
  \bibfield  {author} {\bibinfo {author} {\bibfnamefont {S.}~\bibnamefont
  {Martinelli}}, \bibinfo {author} {\bibfnamefont {H.}~\bibnamefont {Kamada}},
  \bibinfo {author} {\bibfnamefont {G.}~\bibnamefont {Orlandini}}, \ and\
  \bibinfo {author} {\bibfnamefont {W.}~\bibnamefont {Gl\"ockle}},\ }\href
  {\doibase 10.1103/PhysRevC.52.1778} {\bibfield  {journal} {\bibinfo
  {journal} {Phys. Rev. C}\ }\textbf {\bibinfo {volume} {52}},\ \bibinfo
  {pages} {1778} (\bibinfo {year} {1995})}\BibitemShut {NoStop}%
\bibitem [{\citenamefont {Efros}\ \emph
  {et~al.}(1997{\natexlab{a}})\citenamefont {Efros}, \citenamefont
  {Leidemann},\ and\ \citenamefont {Orlandini}}]{efros1997a}%
  \BibitemOpen
  \bibfield  {author} {\bibinfo {author} {\bibfnamefont {V.~D.}\ \bibnamefont
  {Efros}}, \bibinfo {author} {\bibfnamefont {W.}~\bibnamefont {Leidemann}}, \
  and\ \bibinfo {author} {\bibfnamefont {G.}~\bibnamefont {Orlandini}},\ }\href
  {\doibase http://dx.doi.org/10.1016/S0370-2693(97)00772-7} {\bibfield
  {journal} {\bibinfo  {journal} {Physics Letters B}\ }\textbf {\bibinfo
  {volume} {408}},\ \bibinfo {pages} {1 } (\bibinfo {year}
  {1997}{\natexlab{a}})}\BibitemShut {NoStop}%
\bibitem [{\citenamefont {Efros}\ \emph
  {et~al.}(1997{\natexlab{b}})\citenamefont {Efros}, \citenamefont
  {Leidemann},\ and\ \citenamefont {Orlandini}}]{efros1997d}%
  \BibitemOpen
  \bibfield  {author} {\bibinfo {author} {\bibfnamefont {V.~D.}\ \bibnamefont
  {Efros}}, \bibinfo {author} {\bibfnamefont {W.}~\bibnamefont {Leidemann}}, \
  and\ \bibinfo {author} {\bibfnamefont {G.}~\bibnamefont {Orlandini}},\ }\href
  {\doibase 10.1103/PhysRevLett.78.4015} {\bibfield  {journal} {\bibinfo
  {journal} {Phys. Rev. Lett.}\ }\textbf {\bibinfo {volume} {78}},\ \bibinfo
  {pages} {4015} (\bibinfo {year} {1997}{\natexlab{b}})}\BibitemShut {NoStop}%
\bibitem [{\citenamefont {Efros}\ \emph
  {et~al.}(1997{\natexlab{c}})\citenamefont {Efros}, \citenamefont
  {Leidemann},\ and\ \citenamefont {Orlandini}}]{efros1997c}%
  \BibitemOpen
  \bibfield  {author} {\bibinfo {author} {\bibfnamefont {V.~D.}\ \bibnamefont
  {Efros}}, \bibinfo {author} {\bibfnamefont {W.}~\bibnamefont {Leidemann}}, \
  and\ \bibinfo {author} {\bibfnamefont {G.}~\bibnamefont {Orlandini}},\ }\href
  {\doibase 10.1103/PhysRevLett.78.432} {\bibfield  {journal} {\bibinfo
  {journal} {Phys. Rev. Lett.}\ }\textbf {\bibinfo {volume} {78}},\ \bibinfo
  {pages} {432} (\bibinfo {year} {1997}{\natexlab{c}})}\BibitemShut {NoStop}%
\bibitem [{\citenamefont {Efros}\ \emph {et~al.}(1998)\citenamefont {Efros},
  \citenamefont {Leidemann},\ and\ \citenamefont {Orlandini}}]{efros1998}%
  \BibitemOpen
  \bibfield  {author} {\bibinfo {author} {\bibfnamefont {V.~D.}\ \bibnamefont
  {Efros}}, \bibinfo {author} {\bibfnamefont {W.}~\bibnamefont {Leidemann}}, \
  and\ \bibinfo {author} {\bibfnamefont {G.}~\bibnamefont {Orlandini}},\ }\href
  {\doibase 10.1103/PhysRevC.58.582} {\bibfield  {journal} {\bibinfo  {journal}
  {Phys. Rev. C}\ }\textbf {\bibinfo {volume} {58}},\ \bibinfo {pages} {582}
  (\bibinfo {year} {1998})}\BibitemShut {NoStop}%
\bibitem [{\citenamefont {Efros}\ \emph {et~al.}(2000)\citenamefont {Efros},
  \citenamefont {Leidemann}, \citenamefont {Orlandini},\ and\ \citenamefont
  {Tomusiak}}]{efros2000}%
  \BibitemOpen
  \bibfield  {author} {\bibinfo {author} {\bibfnamefont {V.~D.}\ \bibnamefont
  {Efros}}, \bibinfo {author} {\bibfnamefont {W.}~\bibnamefont {Leidemann}},
  \bibinfo {author} {\bibfnamefont {G.}~\bibnamefont {Orlandini}}, \ and\
  \bibinfo {author} {\bibfnamefont {E.~L.}\ \bibnamefont {Tomusiak}},\ }\href
  {\doibase http://dx.doi.org/10.1016/S0370-2693(00)00656-0} {\bibfield
  {journal} {\bibinfo  {journal} {Physics Letters B}\ }\textbf {\bibinfo
  {volume} {484}},\ \bibinfo {pages} {223 } (\bibinfo {year}
  {2000})}\BibitemShut {NoStop}%
\bibitem [{\citenamefont {Bacca}\ \emph {et~al.}(2002)\citenamefont {Bacca},
  \citenamefont {Marchisio}, \citenamefont {Barnea}, \citenamefont
  {Leidemann},\ and\ \citenamefont {Orlandini}}]{bacca2002}%
  \BibitemOpen
  \bibfield  {author} {\bibinfo {author} {\bibfnamefont {S.}~\bibnamefont
  {Bacca}}, \bibinfo {author} {\bibfnamefont {M.~A.}\ \bibnamefont
  {Marchisio}}, \bibinfo {author} {\bibfnamefont {N.}~\bibnamefont {Barnea}},
  \bibinfo {author} {\bibfnamefont {W.}~\bibnamefont {Leidemann}}, \ and\
  \bibinfo {author} {\bibfnamefont {G.}~\bibnamefont {Orlandini}},\ }\href
  {\doibase 10.1103/PhysRevLett.89.052502} {\bibfield  {journal} {\bibinfo
  {journal} {Phys. Rev. Lett.}\ }\textbf {\bibinfo {volume} {89}},\ \bibinfo
  {pages} {052502} (\bibinfo {year} {2002})}\BibitemShut {NoStop}%
\bibitem [{\citenamefont {Bacca}\ \emph
  {et~al.}(2004{\natexlab{a}})\citenamefont {Bacca}, \citenamefont
  {Arenhövel}, \citenamefont {Barnea}, \citenamefont {Leidemann},\ and\
  \citenamefont {Orlandini}}]{bacca2004a}%
  \BibitemOpen
  \bibfield  {author} {\bibinfo {author} {\bibfnamefont {S.}~\bibnamefont
  {Bacca}}, \bibinfo {author} {\bibfnamefont {H.}~\bibnamefont {Arenhövel}},
  \bibinfo {author} {\bibfnamefont {N.}~\bibnamefont {Barnea}}, \bibinfo
  {author} {\bibfnamefont {W.}~\bibnamefont {Leidemann}}, \ and\ \bibinfo
  {author} {\bibfnamefont {G.}~\bibnamefont {Orlandini}},\ }\href {\doibase
  http://dx.doi.org/10.1016/j.physletb.2004.10.025} {\bibfield  {journal}
  {\bibinfo  {journal} {Physics Letters B}\ }\textbf {\bibinfo {volume}
  {603}},\ \bibinfo {pages} {159 } (\bibinfo {year}
  {2004}{\natexlab{a}})}\BibitemShut {NoStop}%
\bibitem [{\citenamefont {Bacca}\ \emph
  {et~al.}(2004{\natexlab{b}})\citenamefont {Bacca}, \citenamefont
  {Arenh{\"o}vel}, \citenamefont {Barnea}, \citenamefont {Leidemann},\ and\
  \citenamefont {Orlandini}}]{BaA04}%
  \BibitemOpen
  \bibfield  {author} {\bibinfo {author} {\bibfnamefont {S.}~\bibnamefont
  {Bacca}}, \bibinfo {author} {\bibfnamefont {H.}~\bibnamefont
  {Arenh{\"o}vel}}, \bibinfo {author} {\bibfnamefont {N.}~\bibnamefont
  {Barnea}}, \bibinfo {author} {\bibfnamefont {W.}~\bibnamefont {Leidemann}}, \
  and\ \bibinfo {author} {\bibfnamefont {G.}~\bibnamefont {Orlandini}},\
  }\href@noop {} {\bibfield  {journal} {\bibinfo  {journal} {Phys.~Lett.~B}\
  }\textbf {\bibinfo {volume} {603}},\ \bibinfo {pages} {159 } (\bibinfo {year}
  {2004}{\natexlab{b}})}\BibitemShut {NoStop}%
\bibitem [{\citenamefont {Gazit}\ \emph
  {et~al.}(2006{\natexlab{a}})\citenamefont {Gazit}, \citenamefont {Bacca},
  \citenamefont {Barnea}, \citenamefont {Leidemann},\ and\ \citenamefont
  {Orlandini}}]{gazit2006}%
  \BibitemOpen
  \bibfield  {author} {\bibinfo {author} {\bibfnamefont {D.}~\bibnamefont
  {Gazit}}, \bibinfo {author} {\bibfnamefont {S.}~\bibnamefont {Bacca}},
  \bibinfo {author} {\bibfnamefont {N.}~\bibnamefont {Barnea}}, \bibinfo
  {author} {\bibfnamefont {W.}~\bibnamefont {Leidemann}}, \ and\ \bibinfo
  {author} {\bibfnamefont {G.}~\bibnamefont {Orlandini}},\ }\href {\doibase
  10.1103/PhysRevLett.96.112301} {\bibfield  {journal} {\bibinfo  {journal}
  {Phys. Rev. Lett.}\ }\textbf {\bibinfo {volume} {96}},\ \bibinfo {pages}
  {112301} (\bibinfo {year} {2006}{\natexlab{a}})}\BibitemShut {NoStop}%
\bibitem [{\citenamefont {Stetcu}\ \emph {et~al.}(2007)\citenamefont {Stetcu},
  \citenamefont {Barrett},\ and\ \citenamefont {van Kolck}}]{stetcu2007}%
  \BibitemOpen
  \bibfield  {author} {\bibinfo {author} {\bibfnamefont {I.}~\bibnamefont
  {Stetcu}}, \bibinfo {author} {\bibfnamefont {B.}~\bibnamefont {Barrett}}, \
  and\ \bibinfo {author} {\bibfnamefont {U.}~\bibnamefont {van Kolck}},\ }\href
  {\doibase 10.1016/j.physletb.2007.07.065} {\bibfield  {journal} {\bibinfo
  {journal} {Physics Letters B}\ }\textbf {\bibinfo {volume} {653}},\ \bibinfo
  {pages} {358 } (\bibinfo {year} {2007})}\BibitemShut {NoStop}%
\bibitem [{\citenamefont {Quaglioni}\ and\ \citenamefont
  {Navrátil}(2007)}]{quaglioni2007}%
  \BibitemOpen
  \bibfield  {author} {\bibinfo {author} {\bibfnamefont {S.}~\bibnamefont
  {Quaglioni}}\ and\ \bibinfo {author} {\bibfnamefont {P.}~\bibnamefont
  {Navrátil}},\ }\href {\doibase
  http://dx.doi.org/10.1016/j.physletb.2007.06.082} {\bibfield  {journal}
  {\bibinfo  {journal} {Physics Letters B}\ }\textbf {\bibinfo {volume}
  {652}},\ \bibinfo {pages} {370 } (\bibinfo {year} {2007})}\BibitemShut
  {NoStop}%
\bibitem [{\citenamefont {Coester}(1958)}]{coester1958}%
  \BibitemOpen
  \bibfield  {author} {\bibinfo {author} {\bibfnamefont {F.}~\bibnamefont
  {Coester}},\ }\href {\doibase 10.1016/0029-5582(58)90280-3} {\bibfield
  {journal} {\bibinfo  {journal} {Nuclear Physics}\ }\textbf {\bibinfo {volume}
  {7}},\ \bibinfo {pages} {421 } (\bibinfo {year} {1958})}\BibitemShut
  {NoStop}%
\bibitem [{\citenamefont {Coester}\ and\ \citenamefont
  {K{\"u}mmel}(1960)}]{coester1960}%
  \BibitemOpen
  \bibfield  {author} {\bibinfo {author} {\bibfnamefont {F.}~\bibnamefont
  {Coester}}\ and\ \bibinfo {author} {\bibfnamefont {H.}~\bibnamefont
  {K{\"u}mmel}},\ }\href {\doibase 10.1016/0029-5582(60)90140-1} {\bibfield
  {journal} {\bibinfo  {journal} {Nuclear Physics}\ }\textbf {\bibinfo {volume}
  {17}},\ \bibinfo {pages} {477 } (\bibinfo {year} {1960})}\BibitemShut
  {NoStop}%
\bibitem [{\citenamefont {{\v {C}}{\'\i}{\v z}ek}(1966)}]{cizek1966}%
  \BibitemOpen
  \bibfield  {author} {\bibinfo {author} {\bibfnamefont {J.}~\bibnamefont {{\v
  {C}}{\'\i}{\v z}ek}},\ }\href {\doibase 10.1063/1.1727484} {\bibfield
  {journal} {\bibinfo  {journal} {The Journal of Chemical Physics}\ }\textbf
  {\bibinfo {volume} {45}},\ \bibinfo {pages} {4256} (\bibinfo {year}
  {1966})}\BibitemShut {NoStop}%
\bibitem [{\citenamefont {{\v {C}}{\'\i}{\v z}ek}(2007)}]{cizek1969}%
  \BibitemOpen
  \bibfield  {author} {\bibinfo {author} {\bibfnamefont {J.}~\bibnamefont {{\v
  {C}}{\'\i}{\v z}ek}},\ }\enquote {\bibinfo {title} {{On the Use of the
  Cluster Expansion and the Technique of Diagrams in Calculations of
  Correlation Effects in Atoms and Molecules}},}\ in\ \href {\doibase
  10.1002/9780470143599.ch2} {\emph {\bibinfo {booktitle} {Advances in Chemical
  Physics}}}\ (\bibinfo  {publisher} {John Wiley \& Sons, Inc.},\ \bibinfo
  {year} {2007})\ pp.\ \bibinfo {pages} {35--89}\BibitemShut {NoStop}%
\bibitem [{\citenamefont {{\v {C}}{\'\i}{\v z}ek}\ and\ \citenamefont
  {Paldus}(1971)}]{cizek1971}%
  \BibitemOpen
  \bibfield  {author} {\bibinfo {author} {\bibfnamefont {J.}~\bibnamefont {{\v
  {C}}{\'\i}{\v z}ek}}\ and\ \bibinfo {author} {\bibfnamefont {J.}~\bibnamefont
  {Paldus}},\ }\href {\doibase 10.1002/qua.560050402} {\bibfield  {journal}
  {\bibinfo  {journal} {International Journal of Quantum Chemistry}\ }\textbf
  {\bibinfo {volume} {5}},\ \bibinfo {pages} {359} (\bibinfo {year}
  {1971})}\BibitemShut {NoStop}%
\bibitem [{\citenamefont {Dean}\ and\ \citenamefont
  {Hjorth-Jensen}(2004)}]{dean2004}%
  \BibitemOpen
  \bibfield  {author} {\bibinfo {author} {\bibfnamefont {D.~J.}\ \bibnamefont
  {Dean}}\ and\ \bibinfo {author} {\bibfnamefont {M.}~\bibnamefont
  {Hjorth-Jensen}},\ }\href {\doibase 10.1103/PhysRevC.69.054320} {\bibfield
  {journal} {\bibinfo  {journal} {Phys. Rev. C}\ }\textbf {\bibinfo {volume}
  {69}},\ \bibinfo {pages} {054320} (\bibinfo {year} {2004})}\BibitemShut
  {NoStop}%
\bibitem [{\citenamefont {Kowalski}\ \emph {et~al.}(2004)\citenamefont
  {Kowalski}, \citenamefont {Dean}, \citenamefont {Hjorth-Jensen},
  \citenamefont {Papenbrock},\ and\ \citenamefont {Piecuch}}]{kowalski2004}%
  \BibitemOpen
  \bibfield  {author} {\bibinfo {author} {\bibfnamefont {K.}~\bibnamefont
  {Kowalski}}, \bibinfo {author} {\bibfnamefont {D.~J.}\ \bibnamefont {Dean}},
  \bibinfo {author} {\bibfnamefont {M.}~\bibnamefont {Hjorth-Jensen}}, \bibinfo
  {author} {\bibfnamefont {T.}~\bibnamefont {Papenbrock}}, \ and\ \bibinfo
  {author} {\bibfnamefont {P.}~\bibnamefont {Piecuch}},\ }\href {\doibase
  10.1103/PhysRevLett.92.132501} {\bibfield  {journal} {\bibinfo  {journal}
  {Phys. Rev. Lett.}\ }\textbf {\bibinfo {volume} {92}},\ \bibinfo {pages}
  {132501} (\bibinfo {year} {2004})}\BibitemShut {NoStop}%
\bibitem [{\citenamefont {Gour}\ \emph {et~al.}(2006)\citenamefont {Gour},
  \citenamefont {Piecuch}, \citenamefont {Hjorth-Jensen}, \citenamefont
  {W\l{}och},\ and\ \citenamefont {Dean}}]{gour2006}%
  \BibitemOpen
  \bibfield  {author} {\bibinfo {author} {\bibfnamefont {J.~R.}\ \bibnamefont
  {Gour}}, \bibinfo {author} {\bibfnamefont {P.}~\bibnamefont {Piecuch}},
  \bibinfo {author} {\bibfnamefont {M.}~\bibnamefont {Hjorth-Jensen}}, \bibinfo
  {author} {\bibfnamefont {M.}~\bibnamefont {W\l{}och}}, \ and\ \bibinfo
  {author} {\bibfnamefont {D.~J.}\ \bibnamefont {Dean}},\ }\href {\doibase
  10.1103/PhysRevC.74.024310} {\bibfield  {journal} {\bibinfo  {journal} {Phys.
  Rev. C}\ }\textbf {\bibinfo {volume} {74}},\ \bibinfo {pages} {024310}
  (\bibinfo {year} {2006})}\BibitemShut {NoStop}%
\bibitem [{\citenamefont {Hagen}\ \emph {et~al.}(2008)\citenamefont {Hagen},
  \citenamefont {Papenbrock}, \citenamefont {Dean},\ and\ \citenamefont
  {Hjorth-Jensen}}]{hagen2008}%
  \BibitemOpen
  \bibfield  {author} {\bibinfo {author} {\bibfnamefont {G.}~\bibnamefont
  {Hagen}}, \bibinfo {author} {\bibfnamefont {T.}~\bibnamefont {Papenbrock}},
  \bibinfo {author} {\bibfnamefont {D.~J.}\ \bibnamefont {Dean}}, \ and\
  \bibinfo {author} {\bibfnamefont {M.}~\bibnamefont {Hjorth-Jensen}},\ }\href
  {\doibase 10.1103/PhysRevLett.101.092502} {\bibfield  {journal} {\bibinfo
  {journal} {Phys. Rev. Lett.}\ }\textbf {\bibinfo {volume} {101}},\ \bibinfo
  {pages} {092502} (\bibinfo {year} {2008})}\BibitemShut {NoStop}%
\bibitem [{\citenamefont {Hagen}\ \emph {et~al.}(2010)\citenamefont {Hagen},
  \citenamefont {Papenbrock}, \citenamefont {Dean},\ and\ \citenamefont
  {Hjorth-Jensen}}]{hagen2010b}%
  \BibitemOpen
  \bibfield  {author} {\bibinfo {author} {\bibfnamefont {G.}~\bibnamefont
  {Hagen}}, \bibinfo {author} {\bibfnamefont {T.}~\bibnamefont {Papenbrock}},
  \bibinfo {author} {\bibfnamefont {D.~J.}\ \bibnamefont {Dean}}, \ and\
  \bibinfo {author} {\bibfnamefont {M.}~\bibnamefont {Hjorth-Jensen}},\ }\href
  {\doibase 10.1103/PhysRevC.82.034330} {\bibfield  {journal} {\bibinfo
  {journal} {Phys. Rev. C}\ }\textbf {\bibinfo {volume} {82}},\ \bibinfo
  {pages} {034330} (\bibinfo {year} {2010})}\BibitemShut {NoStop}%
\bibitem [{\citenamefont {Hagen}\ \emph
  {et~al.}(2012{\natexlab{b}})\citenamefont {Hagen}, \citenamefont
  {Hjorth-Jensen}, \citenamefont {Jansen}, \citenamefont {Machleidt},\ and\
  \citenamefont {Papenbrock}}]{hagen2012a}%
  \BibitemOpen
  \bibfield  {author} {\bibinfo {author} {\bibfnamefont {G.}~\bibnamefont
  {Hagen}}, \bibinfo {author} {\bibfnamefont {M.}~\bibnamefont
  {Hjorth-Jensen}}, \bibinfo {author} {\bibfnamefont {G.~R.}\ \bibnamefont
  {Jansen}}, \bibinfo {author} {\bibfnamefont {R.}~\bibnamefont {Machleidt}}, \
  and\ \bibinfo {author} {\bibfnamefont {T.}~\bibnamefont {Papenbrock}},\
  }\href {\doibase 10.1103/PhysRevLett.108.242501} {\bibfield  {journal}
  {\bibinfo  {journal} {Phys. Rev. Lett.}\ }\textbf {\bibinfo {volume} {108}},\
  \bibinfo {pages} {242501} (\bibinfo {year} {2012}{\natexlab{b}})}\BibitemShut
  {NoStop}%
\bibitem [{\citenamefont {Roth}\ \emph {et~al.}(2012)\citenamefont {Roth},
  \citenamefont {Binder}, \citenamefont {Vobig}, \citenamefont {Calci},
  \citenamefont {Langhammer},\ and\ \citenamefont {Navr\'atil}}]{roth2012}%
  \BibitemOpen
  \bibfield  {author} {\bibinfo {author} {\bibfnamefont {R.}~\bibnamefont
  {Roth}}, \bibinfo {author} {\bibfnamefont {S.}~\bibnamefont {Binder}},
  \bibinfo {author} {\bibfnamefont {K.}~\bibnamefont {Vobig}}, \bibinfo
  {author} {\bibfnamefont {A.}~\bibnamefont {Calci}}, \bibinfo {author}
  {\bibfnamefont {J.}~\bibnamefont {Langhammer}}, \ and\ \bibinfo {author}
  {\bibfnamefont {P.}~\bibnamefont {Navr\'atil}},\ }\href {\doibase
  10.1103/PhysRevLett.109.052501} {\bibfield  {journal} {\bibinfo  {journal}
  {Phys. Rev. Lett.}\ }\textbf {\bibinfo {volume} {109}},\ \bibinfo {pages}
  {052501} (\bibinfo {year} {2012})}\BibitemShut {NoStop}%
\bibitem [{\citenamefont {Binder}\ \emph {et~al.}(2013)\citenamefont {Binder},
  \citenamefont {Piecuch}, \citenamefont {Calci}, \citenamefont {Langhammer},
  \citenamefont {Navr\'atil},\ and\ \citenamefont {Roth}}]{binder2013}%
  \BibitemOpen
  \bibfield  {author} {\bibinfo {author} {\bibfnamefont {S.}~\bibnamefont
  {Binder}}, \bibinfo {author} {\bibfnamefont {P.}~\bibnamefont {Piecuch}},
  \bibinfo {author} {\bibfnamefont {A.}~\bibnamefont {Calci}}, \bibinfo
  {author} {\bibfnamefont {J.}~\bibnamefont {Langhammer}}, \bibinfo {author}
  {\bibfnamefont {P.}~\bibnamefont {Navr\'atil}}, \ and\ \bibinfo {author}
  {\bibfnamefont {R.}~\bibnamefont {Roth}},\ }\href {\doibase
  10.1103/PhysRevC.88.054319} {\bibfield  {journal} {\bibinfo  {journal} {Phys.
  Rev. C}\ }\textbf {\bibinfo {volume} {88}},\ \bibinfo {pages} {054319}
  (\bibinfo {year} {2013})}\BibitemShut {NoStop}%
\bibitem [{\citenamefont {Binder}\ \emph {et~al.}(2014)\citenamefont {Binder},
  \citenamefont {Langhammer}, \citenamefont {Calci},\ and\ \citenamefont
  {Roth}}]{binder2014}%
  \BibitemOpen
  \bibfield  {author} {\bibinfo {author} {\bibfnamefont {S.}~\bibnamefont
  {Binder}}, \bibinfo {author} {\bibfnamefont {J.}~\bibnamefont {Langhammer}},
  \bibinfo {author} {\bibfnamefont {A.}~\bibnamefont {Calci}}, \ and\ \bibinfo
  {author} {\bibfnamefont {R.}~\bibnamefont {Roth}},\ }\href {\doibase
  http://dx.doi.org/10.1016/j.physletb.2014.07.010} {\bibfield  {journal}
  {\bibinfo  {journal} {Physics Letters B}\ }\textbf {\bibinfo {volume}
  {736}},\ \bibinfo {pages} {119 } (\bibinfo {year} {2014})}\BibitemShut
  {NoStop}%
\bibitem [{\citenamefont {Hagen}\ \emph {et~al.}(2014)\citenamefont {Hagen},
  \citenamefont {Papenbrock}, \citenamefont {Hjorth-Jensen},\ and\
  \citenamefont {Dean}}]{hagen2013c}%
  \BibitemOpen
  \bibfield  {author} {\bibinfo {author} {\bibfnamefont {G.}~\bibnamefont
  {Hagen}}, \bibinfo {author} {\bibfnamefont {T.}~\bibnamefont {Papenbrock}},
  \bibinfo {author} {\bibfnamefont {M.}~\bibnamefont {Hjorth-Jensen}}, \ and\
  \bibinfo {author} {\bibfnamefont {D.~J.}\ \bibnamefont {Dean}},\ }\href
  {http://stacks.iop.org/0034-4885/77/i=9/a=096302} {\bibfield  {journal}
  {\bibinfo  {journal} {Reports on Progress in Physics}\ }\textbf {\bibinfo
  {volume} {77}},\ \bibinfo {pages} {096302} (\bibinfo {year}
  {2014})}\BibitemShut {NoStop}%
\bibitem [{\citenamefont {Bacca}\ \emph {et~al.}(2013)\citenamefont {Bacca},
  \citenamefont {Barnea}, \citenamefont {Hagen}, \citenamefont {Orlandini},\
  and\ \citenamefont {Papenbrock}}]{bacca2013}%
  \BibitemOpen
  \bibfield  {author} {\bibinfo {author} {\bibfnamefont {S.}~\bibnamefont
  {Bacca}}, \bibinfo {author} {\bibfnamefont {N.}~\bibnamefont {Barnea}},
  \bibinfo {author} {\bibfnamefont {G.}~\bibnamefont {Hagen}}, \bibinfo
  {author} {\bibfnamefont {G.}~\bibnamefont {Orlandini}}, \ and\ \bibinfo
  {author} {\bibfnamefont {T.}~\bibnamefont {Papenbrock}},\ }\href {\doibase
  10.1103/PhysRevLett.111.122502} {\bibfield  {journal} {\bibinfo  {journal}
  {Phys. Rev. Lett.}\ }\textbf {\bibinfo {volume} {111}},\ \bibinfo {pages}
  {122502} (\bibinfo {year} {2013})}\BibitemShut {NoStop}%
\bibitem [{\citenamefont {Efros}\ \emph {et~al.}(1999)\citenamefont {Efros},
  \citenamefont {Leidemann},\ and\ \citenamefont {Orlandini}}]{efros1999}%
  \BibitemOpen
  \bibfield  {author} {\bibinfo {author} {\bibfnamefont {V.~D.}\ \bibnamefont
  {Efros}}, \bibinfo {author} {\bibfnamefont {W.}~\bibnamefont {Leidemann}}, \
  and\ \bibinfo {author} {\bibfnamefont {G.}~\bibnamefont {Orlandini}},\ }\href
  {\doibase 10.1007/s006010050118} {\bibfield  {journal} {\bibinfo  {journal}
  {Few-Body Systems}\ }\textbf {\bibinfo {volume} {26}},\ \bibinfo {pages}
  {251} (\bibinfo {year} {1999})}\BibitemShut {NoStop}%
\bibitem [{\citenamefont {{Andreasi, D.}}\ \emph {et~al.}(2005)\citenamefont
  {{Andreasi, D.}}, \citenamefont {{Leidemann, W.}}, \citenamefont {{Reiß,
  C.}},\ and\ \citenamefont {{Schwamb, M.}}}]{andreasi2005}%
  \BibitemOpen
  \bibfield  {author} {\bibinfo {author} {\bibnamefont {{Andreasi, D.}}},
  \bibinfo {author} {\bibnamefont {{Leidemann, W.}}}, \bibinfo {author}
  {\bibnamefont {{Reiß, C.}}}, \ and\ \bibinfo {author} {\bibnamefont
  {{Schwamb, M.}}},\ }\href {\doibase 10.1140/epja/i2005-10009-3} {\bibfield
  {journal} {\bibinfo  {journal} {Eur. Phys. J. A}\ }\textbf {\bibinfo {volume}
  {24}},\ \bibinfo {pages} {361} (\bibinfo {year} {2005})}\BibitemShut
  {NoStop}%
\bibitem [{\citenamefont {Piana}\ and\ \citenamefont
  {Leidemann}(2000)}]{lapiana2000}%
  \BibitemOpen
  \bibfield  {author} {\bibinfo {author} {\bibfnamefont {A.~L.}\ \bibnamefont
  {Piana}}\ and\ \bibinfo {author} {\bibfnamefont {W.}~\bibnamefont
  {Leidemann}},\ }\href {\doibase
  http://dx.doi.org/10.1016/S0375-9474(00)00310-9} {\bibfield  {journal}
  {\bibinfo  {journal} {Nuclear Physics A}\ }\textbf {\bibinfo {volume}
  {677}},\ \bibinfo {pages} {423 } (\bibinfo {year} {2000})}\BibitemShut
  {NoStop}%
\bibitem [{\citenamefont {Golak}\ \emph {et~al.}(2002)\citenamefont {Golak},
  \citenamefont {Skibiński}, \citenamefont {Glöckle}, \citenamefont {Kamada},
  \citenamefont {Nogga}, \citenamefont {Witała}, \citenamefont {Efros},
  \citenamefont {Leidemann}, \citenamefont {Orlandini},\ and\ \citenamefont
  {Tomusiak}}]{golak2002}%
  \BibitemOpen
  \bibfield  {author} {\bibinfo {author} {\bibfnamefont {J.}~\bibnamefont
  {Golak}}, \bibinfo {author} {\bibfnamefont {R.}~\bibnamefont {Skibiński}},
  \bibinfo {author} {\bibfnamefont {W.}~\bibnamefont {Glöckle}}, \bibinfo
  {author} {\bibfnamefont {H.}~\bibnamefont {Kamada}}, \bibinfo {author}
  {\bibfnamefont {A.}~\bibnamefont {Nogga}}, \bibinfo {author} {\bibfnamefont
  {H.}~\bibnamefont {Witała}}, \bibinfo {author} {\bibfnamefont
  {V.}~\bibnamefont {Efros}}, \bibinfo {author} {\bibfnamefont
  {W.}~\bibnamefont {Leidemann}}, \bibinfo {author} {\bibfnamefont
  {G.}~\bibnamefont {Orlandini}}, \ and\ \bibinfo {author} {\bibfnamefont
  {E.}~\bibnamefont {Tomusiak}},\ }\href {\doibase
  http://dx.doi.org/10.1016/S0375-9474(02)00989-2} {\bibfield  {journal}
  {\bibinfo  {journal} {Nuclear Physics A}\ }\textbf {\bibinfo {volume}
  {707}},\ \bibinfo {pages} {365 } (\bibinfo {year} {2002})}\BibitemShut
  {NoStop}%
\bibitem [{\citenamefont {Efros}\ \emph
  {et~al.}(2007{\natexlab{a}})\citenamefont {Efros}, \citenamefont {Leidemann},
  \citenamefont {Orlandini},\ and\ \citenamefont {Barnea}}]{efros2007b}%
  \BibitemOpen
  \bibfield  {author} {\bibinfo {author} {\bibfnamefont {V.~D.}\ \bibnamefont
  {Efros}}, \bibinfo {author} {\bibfnamefont {W.}~\bibnamefont {Leidemann}},
  \bibinfo {author} {\bibfnamefont {G.}~\bibnamefont {Orlandini}}, \ and\
  \bibinfo {author} {\bibfnamefont {N.}~\bibnamefont {Barnea}},\ }\href
  {http://stacks.iop.org/0954-3899/34/i=12/a=R02} {\bibfield  {journal}
  {\bibinfo  {journal} {Journal of Physics G: Nuclear and Particle Physics}\
  }\textbf {\bibinfo {volume} {34}},\ \bibinfo {pages} {R459} (\bibinfo {year}
  {2007}{\natexlab{a}})}\BibitemShut {NoStop}%
\bibitem [{\citenamefont {{Bacca}}\ and\ \citenamefont
  {{Pastore}}(2014)}]{bacca2014}%
  \BibitemOpen
  \bibfield  {author} {\bibinfo {author} {\bibfnamefont {S.}~\bibnamefont
  {{Bacca}}}\ and\ \bibinfo {author} {\bibfnamefont {S.}~\bibnamefont
  {{Pastore}}},\ }\href {http://adsabs.harvard.edu/abs/2014arXiv1407.3490B}
  {\bibfield  {journal} {\bibinfo  {journal} {ArXiv e-prints}\ } (\bibinfo
  {year} {2014})},\ \Eprint {http://arxiv.org/abs/1407.3490} {arXiv:1407.3490
  [nucl-th]} \BibitemShut {NoStop}%
\bibitem [{\citenamefont {Bartlett}\ and\ \citenamefont
  {Musia\l{}}(2007)}]{bartlett2007}%
  \BibitemOpen
  \bibfield  {author} {\bibinfo {author} {\bibfnamefont {R.~J.}\ \bibnamefont
  {Bartlett}}\ and\ \bibinfo {author} {\bibfnamefont {M.}~\bibnamefont
  {Musia\l{}}},\ }\href {\doibase 10.1103/RevModPhys.79.291} {\bibfield
  {journal} {\bibinfo  {journal} {Rev. Mod. Phys.}\ }\textbf {\bibinfo {volume}
  {79}},\ \bibinfo {pages} {291} (\bibinfo {year} {2007})}\BibitemShut
  {NoStop}%
\bibitem [{\citenamefont {Shavitt}\ and\ \citenamefont
  {Bartlett}(2009)}]{shavittbartlett2009}%
  \BibitemOpen
  \bibfield  {author} {\bibinfo {author} {\bibfnamefont {I.}~\bibnamefont
  {Shavitt}}\ and\ \bibinfo {author} {\bibfnamefont {R.~J.}\ \bibnamefont
  {Bartlett}},\ }\href@noop {} {\emph {\bibinfo {title} {Many-body Methods in
  Chemistry and Physics}}}\ (\bibinfo  {publisher} {Cambridge University
  Press},\ \bibinfo {year} {2009})\BibitemShut {NoStop}%
\bibitem [{\citenamefont {Stanton}\ and\ \citenamefont
  {Bartlett}(1993)}]{stanton1993}%
  \BibitemOpen
  \bibfield  {author} {\bibinfo {author} {\bibfnamefont {J.~F.}\ \bibnamefont
  {Stanton}}\ and\ \bibinfo {author} {\bibfnamefont {R.~J.}\ \bibnamefont
  {Bartlett}},\ }\href {\doibase 10.1063/1.464746} {\bibfield  {journal}
  {\bibinfo  {journal} {The Journal of Chemical Physics}\ }\textbf {\bibinfo
  {volume} {98}},\ \bibinfo {pages} {7029} (\bibinfo {year}
  {1993})}\BibitemShut {NoStop}%
\bibitem [{\citenamefont {Efros}\ \emph
  {et~al.}(2007{\natexlab{b}})\citenamefont {Efros}, \citenamefont {Leidemann},
  \citenamefont {Orlandini},\ and\ \citenamefont {Barnea}}]{efros2007}%
  \BibitemOpen
  \bibfield  {author} {\bibinfo {author} {\bibfnamefont {V.~D.}\ \bibnamefont
  {Efros}}, \bibinfo {author} {\bibfnamefont {W.}~\bibnamefont {Leidemann}},
  \bibinfo {author} {\bibfnamefont {G.}~\bibnamefont {Orlandini}}, \ and\
  \bibinfo {author} {\bibfnamefont {N.}~\bibnamefont {Barnea}},\ }\href
  {http://stacks.iop.org/0954-3899/34/i=12/a=R02} {\bibfield  {journal}
  {\bibinfo  {journal} {Journal of Physics G: Nuclear and Particle Physics}\
  }\textbf {\bibinfo {volume} {34}},\ \bibinfo {pages} {R459} (\bibinfo {year}
  {2007}{\natexlab{b}})}\BibitemShut {NoStop}%
\bibitem [{\citenamefont {Collum}(1995)}]{Col95}%
  \BibitemOpen
  \bibfield  {author} {\bibinfo {author} {\bibfnamefont {J.}~\bibnamefont
  {Collum}},\ }\href@noop {} {\bibfield  {journal} {\bibinfo  {journal}
  {Technical Report, University of Maryland}\ }\textbf {\bibinfo {volume}
  {TR-3576}} (\bibinfo {year} {1995})}\BibitemShut {NoStop}%
\bibitem [{\citenamefont {Hagen}\ \emph {et~al.}(2009)\citenamefont {Hagen},
  \citenamefont {Papenbrock},\ and\ \citenamefont {Dean}}]{hagen2009a}%
  \BibitemOpen
  \bibfield  {author} {\bibinfo {author} {\bibfnamefont {G.}~\bibnamefont
  {Hagen}}, \bibinfo {author} {\bibfnamefont {T.}~\bibnamefont {Papenbrock}}, \
  and\ \bibinfo {author} {\bibfnamefont {D.~J.}\ \bibnamefont {Dean}},\ }\href
  {\doibase 10.1103/PhysRevLett.103.062503} {\bibfield  {journal} {\bibinfo
  {journal} {Phys. Rev. Lett.}\ }\textbf {\bibinfo {volume} {103}},\ \bibinfo
  {pages} {062503} (\bibinfo {year} {2009})}\BibitemShut {NoStop}%
\bibitem [{\citenamefont {Jansen}(2013)}]{jansen2012}%
  \BibitemOpen
  \bibfield  {author} {\bibinfo {author} {\bibfnamefont {G.~R.}\ \bibnamefont
  {Jansen}},\ }\href {\doibase 10.1103/PhysRevC.88.024305} {\bibfield
  {journal} {\bibinfo  {journal} {Phys. Rev. C}\ }\textbf {\bibinfo {volume}
  {88}},\ \bibinfo {pages} {024305} (\bibinfo {year} {2013})}\BibitemShut
  {NoStop}%
\bibitem [{\citenamefont {Bacca}\ \emph {et~al.}(2007)\citenamefont {Bacca},
  \citenamefont {Arenh\"ovel}, \citenamefont {Barnea}, \citenamefont
  {Leidemann},\ and\ \citenamefont {Orlandini}}]{bacca2007}%
  \BibitemOpen
  \bibfield  {author} {\bibinfo {author} {\bibfnamefont {S.}~\bibnamefont
  {Bacca}}, \bibinfo {author} {\bibfnamefont {H.}~\bibnamefont {Arenh\"ovel}},
  \bibinfo {author} {\bibfnamefont {N.}~\bibnamefont {Barnea}}, \bibinfo
  {author} {\bibfnamefont {W.}~\bibnamefont {Leidemann}}, \ and\ \bibinfo
  {author} {\bibfnamefont {G.}~\bibnamefont {Orlandini}},\ }\href {\doibase
  10.1103/PhysRevC.76.014003} {\bibfield  {journal} {\bibinfo  {journal} {Phys.
  Rev. C}\ }\textbf {\bibinfo {volume} {76}},\ \bibinfo {pages} {014003}
  (\bibinfo {year} {2007})}\BibitemShut {NoStop}%
\bibitem [{\citenamefont {Barnea}\ \emph {et~al.}(2001)\citenamefont {Barnea},
  \citenamefont {Leidemann},\ and\ \citenamefont {Orlandini}}]{barnea2001}%
  \BibitemOpen
  \bibfield  {author} {\bibinfo {author} {\bibfnamefont {N.}~\bibnamefont
  {Barnea}}, \bibinfo {author} {\bibfnamefont {W.}~\bibnamefont {Leidemann}}, \
  and\ \bibinfo {author} {\bibfnamefont {G.}~\bibnamefont {Orlandini}},\ }\href
  {\doibase http://dx.doi.org/10.1016/S0375-9474(01)00794-1} {\bibfield
  {journal} {\bibinfo  {journal} {Nuclear Physics A}\ }\textbf {\bibinfo
  {volume} {693}},\ \bibinfo {pages} {565 } (\bibinfo {year}
  {2001})}\BibitemShut {NoStop}%
\bibitem [{\citenamefont {Entem}\ and\ \citenamefont
  {Machleidt}(2003)}]{entem2003}%
  \BibitemOpen
  \bibfield  {author} {\bibinfo {author} {\bibfnamefont {D.~R.}\ \bibnamefont
  {Entem}}\ and\ \bibinfo {author} {\bibfnamefont {R.}~\bibnamefont
  {Machleidt}},\ }\href {\doibase 10.1103/PhysRevC.68.041001} {\bibfield
  {journal} {\bibinfo  {journal} {Phys. Rev. C}\ }\textbf {\bibinfo {volume}
  {68}},\ \bibinfo {pages} {041001} (\bibinfo {year} {2003})}\BibitemShut
  {NoStop}%
\bibitem [{\citenamefont {Arkatov}\ \emph {et~al.}(1979)\citenamefont {Arkatov}
  \emph {et~al.}}]{arkatov}%
  \BibitemOpen
  \bibfield  {author} {\bibinfo {author} {\bibfnamefont {Y.~M.}\ \bibnamefont
  {Arkatov}} \emph {et~al.},\ }\href@noop {} {\bibfield  {journal} {\bibinfo
  {journal} {Yad. Konst.}\ }\textbf {\bibinfo {volume} {4}},\ \bibinfo {pages}
  {55} (\bibinfo {year} {1979})}\BibitemShut {NoStop}%
\bibitem [{\citenamefont {Nilsson}\ \emph {et~al.}(2005)\citenamefont
  {Nilsson}, \citenamefont {Adler}, \citenamefont {Andersson}, \citenamefont
  {Annand}, \citenamefont {Akkurt}, \citenamefont {Boland}, \citenamefont
  {Crawford}, \citenamefont {Fissum}, \citenamefont {Hansen}, \citenamefont
  {Harty}, \citenamefont {Ireland}, \citenamefont {Isaksson}, \citenamefont
  {Karlsson}, \citenamefont {Lundin}, \citenamefont {McGeorge}, \citenamefont
  {Miller}, \citenamefont {Ruijter}, \citenamefont {Sandell}, \citenamefont
  {Schr{\"o}der}, \citenamefont {Sims},\ and\ \citenamefont
  {Watts}}]{nilsson2005}%
  \BibitemOpen
  \bibfield  {author} {\bibinfo {author} {\bibfnamefont {B.}~\bibnamefont
  {Nilsson}}, \bibinfo {author} {\bibfnamefont {J.-O.}\ \bibnamefont {Adler}},
  \bibinfo {author} {\bibfnamefont {B.-E.}\ \bibnamefont {Andersson}}, \bibinfo
  {author} {\bibfnamefont {J.}~\bibnamefont {Annand}}, \bibinfo {author}
  {\bibfnamefont {I.}~\bibnamefont {Akkurt}}, \bibinfo {author} {\bibfnamefont
  {M.}~\bibnamefont {Boland}}, \bibinfo {author} {\bibfnamefont
  {G.}~\bibnamefont {Crawford}}, \bibinfo {author} {\bibfnamefont
  {K.}~\bibnamefont {Fissum}}, \bibinfo {author} {\bibfnamefont
  {K.}~\bibnamefont {Hansen}}, \bibinfo {author} {\bibfnamefont
  {P.}~\bibnamefont {Harty}}, \bibinfo {author} {\bibfnamefont
  {D.}~\bibnamefont {Ireland}}, \bibinfo {author} {\bibfnamefont
  {L.}~\bibnamefont {Isaksson}}, \bibinfo {author} {\bibfnamefont
  {M.}~\bibnamefont {Karlsson}}, \bibinfo {author} {\bibfnamefont
  {M.}~\bibnamefont {Lundin}}, \bibinfo {author} {\bibfnamefont
  {J.}~\bibnamefont {McGeorge}}, \bibinfo {author} {\bibfnamefont
  {G.}~\bibnamefont {Miller}}, \bibinfo {author} {\bibfnamefont
  {H.}~\bibnamefont {Ruijter}}, \bibinfo {author} {\bibfnamefont
  {A.}~\bibnamefont {Sandell}}, \bibinfo {author} {\bibfnamefont
  {B.}~\bibnamefont {Schr{\"o}der}}, \bibinfo {author} {\bibfnamefont
  {D.}~\bibnamefont {Sims}}, \ and\ \bibinfo {author} {\bibfnamefont
  {D.}~\bibnamefont {Watts}},\ }\href {\doibase
  http://dx.doi.org/10.1016/j.physletb.2005.08.081} {\bibfield  {journal}
  {\bibinfo  {journal} {Physics Letters B}\ }\textbf {\bibinfo {volume}
  {626}},\ \bibinfo {pages} {65 } (\bibinfo {year} {2005})}\BibitemShut
  {NoStop}%
\bibitem [{\citenamefont {Raut}\ \emph {et~al.}(2012)\citenamefont {Raut},
  \citenamefont {Tornow}, \citenamefont {Ahmed}, \citenamefont {Crowell},
  \citenamefont {Kelley}, \citenamefont {Rusev}, \citenamefont {Stave},\ and\
  \citenamefont {Tonchev}}]{raut2012}%
  \BibitemOpen
  \bibfield  {author} {\bibinfo {author} {\bibfnamefont {R.}~\bibnamefont
  {Raut}}, \bibinfo {author} {\bibfnamefont {W.}~\bibnamefont {Tornow}},
  \bibinfo {author} {\bibfnamefont {M.~W.}\ \bibnamefont {Ahmed}}, \bibinfo
  {author} {\bibfnamefont {A.~S.}\ \bibnamefont {Crowell}}, \bibinfo {author}
  {\bibfnamefont {J.~H.}\ \bibnamefont {Kelley}}, \bibinfo {author}
  {\bibfnamefont {G.}~\bibnamefont {Rusev}}, \bibinfo {author} {\bibfnamefont
  {S.~C.}\ \bibnamefont {Stave}}, \ and\ \bibinfo {author} {\bibfnamefont
  {A.~P.}\ \bibnamefont {Tonchev}},\ }\href {\doibase
  10.1103/PhysRevLett.108.042502} {\bibfield  {journal} {\bibinfo  {journal}
  {Phys. Rev. Lett.}\ }\textbf {\bibinfo {volume} {108}},\ \bibinfo {pages}
  {042502} (\bibinfo {year} {2012})}\BibitemShut {NoStop}%
\bibitem [{\citenamefont {Ahrens}\ \emph {et~al.}(1975)\citenamefont {Ahrens},
  \citenamefont {Borchert}, \citenamefont {Czock}, \citenamefont {Eppler},
  \citenamefont {Gimm}, \citenamefont {Gundrum}, \citenamefont {Kr{\"o}ning},
  \citenamefont {Riehn}, \citenamefont {Ram}, \citenamefont {Zieger},\ and\
  \citenamefont {Ziegler}}]{ahrens1975}%
  \BibitemOpen
  \bibfield  {author} {\bibinfo {author} {\bibfnamefont {J.}~\bibnamefont
  {Ahrens}}, \bibinfo {author} {\bibfnamefont {H.}~\bibnamefont {Borchert}},
  \bibinfo {author} {\bibfnamefont {K.}~\bibnamefont {Czock}}, \bibinfo
  {author} {\bibfnamefont {H.}~\bibnamefont {Eppler}}, \bibinfo {author}
  {\bibfnamefont {H.}~\bibnamefont {Gimm}}, \bibinfo {author} {\bibfnamefont
  {H.}~\bibnamefont {Gundrum}}, \bibinfo {author} {\bibfnamefont
  {M.}~\bibnamefont {Kr{\"o}ning}}, \bibinfo {author} {\bibfnamefont
  {P.}~\bibnamefont {Riehn}}, \bibinfo {author} {\bibfnamefont {G.~S.}\
  \bibnamefont {Ram}}, \bibinfo {author} {\bibfnamefont {A.}~\bibnamefont
  {Zieger}}, \ and\ \bibinfo {author} {\bibfnamefont {B.}~\bibnamefont
  {Ziegler}},\ }\href {\doibase 10.1016/0375-9474(75)90543-6} {\bibfield
  {journal} {\bibinfo  {journal} {Nuclear Physics A}\ }\textbf {\bibinfo
  {volume} {251}},\ \bibinfo {pages} {479 } (\bibinfo {year}
  {1975})}\BibitemShut {NoStop}%
\bibitem [{\citenamefont {Taube}\ and\ \citenamefont
  {Bartlett}(2008)}]{taube2008}%
  \BibitemOpen
  \bibfield  {author} {\bibinfo {author} {\bibfnamefont {A.~G.}\ \bibnamefont
  {Taube}}\ and\ \bibinfo {author} {\bibfnamefont {R.~J.}\ \bibnamefont
  {Bartlett}},\ }\href {\doibase 10.1063/1.2830236} {\bibfield  {journal}
  {\bibinfo  {journal} {The Journal of Chemical Physics}\ }\textbf {\bibinfo
  {volume} {128}},\ \bibinfo {eid} {044110} (\bibinfo {year}
  {2008})}\BibitemShut {NoStop}%
\bibitem [{\citenamefont {Ishkhanov}\ \emph {et~al.}(2002)\citenamefont
  {Ishkhanov}, \citenamefont {Kapitonov}, \citenamefont {Lileeva},
  \citenamefont {Shirokov}, \citenamefont {Erokhova}, \citenamefont {Elkin},\
  and\ \citenamefont {Izotova}}]{ishkhanov2002}%
  \BibitemOpen
  \bibfield  {author} {\bibinfo {author} {\bibfnamefont {B.~S.}\ \bibnamefont
  {Ishkhanov}}, \bibinfo {author} {\bibfnamefont {I.~M.}\ \bibnamefont
  {Kapitonov}}, \bibinfo {author} {\bibfnamefont {E.~I.}\ \bibnamefont
  {Lileeva}}, \bibinfo {author} {\bibfnamefont {E.~V.}\ \bibnamefont
  {Shirokov}}, \bibinfo {author} {\bibfnamefont {V.~A.}\ \bibnamefont
  {Erokhova}}, \bibinfo {author} {\bibfnamefont {M.~A.}\ \bibnamefont {Elkin}},
  \ and\ \bibinfo {author} {\bibfnamefont {A.~V.}\ \bibnamefont {Izotova}},\
  }\href@noop {} {\emph {\bibinfo {title} {Cross sections of photon absorption
  by nuclei with nucleon numbers 12 - 65}}},\ \bibinfo {type} {Tech. Rep.}\
  \bibinfo {number} {MSU-INP-2002-27/711}\ (\bibinfo  {institution} {Institute
  of Nuclear Physics},\ \bibinfo {address} {Moscow State University},\ \bibinfo
  {year} {2002})\BibitemShut {NoStop}%
\bibitem [{\citenamefont {Ishkhanov}\ and\ \citenamefont
  {Orlin}(2004)}]{ishkhanov2004}%
  \BibitemOpen
  \bibfield  {author} {\bibinfo {author} {\bibfnamefont {B.}~\bibnamefont
  {Ishkhanov}}\ and\ \bibinfo {author} {\bibfnamefont {V.}~\bibnamefont
  {Orlin}},\ }\href {\doibase 10.1134/1.1755384} {\bibfield  {journal}
  {\bibinfo  {journal} {Physics of Atomic Nuclei}\ }\textbf {\bibinfo {volume}
  {67}},\ \bibinfo {pages} {920} (\bibinfo {year} {2004})}\BibitemShut
  {NoStop}%
\bibitem [{\citenamefont {Orlandini}\ and\ \citenamefont
  {Traini}(1991)}]{Orlandini91}%
  \BibitemOpen
  \bibfield  {author} {\bibinfo {author} {\bibfnamefont {G.}~\bibnamefont
  {Orlandini}}\ and\ \bibinfo {author} {\bibfnamefont {M.}~\bibnamefont
  {Traini}},\ }\href {http://stacks.iop.org/0034-4885/54/i=2/a=002} {\bibfield
  {journal} {\bibinfo  {journal} {Reports on Progress in Physics}\ }\textbf
  {\bibinfo {volume} {54}},\ \bibinfo {pages} {257} (\bibinfo {year}
  {1991})}\BibitemShut {NoStop}%
\bibitem [{\citenamefont {Brink}(1957)}]{brink1957}%
  \BibitemOpen
  \bibfield  {author} {\bibinfo {author} {\bibfnamefont {D.}~\bibnamefont
  {Brink}},\ }\href {\doibase http://dx.doi.org/10.1016/0029-5582(87)90021-6}
  {\bibfield  {journal} {\bibinfo  {journal} {Nuclear Physics}\ }\textbf
  {\bibinfo {volume} {4}},\ \bibinfo {pages} {215 } (\bibinfo {year}
  {1957})}\BibitemShut {NoStop}%
\bibitem [{\citenamefont {Repko}\ \emph {et~al.}(2013)\citenamefont {Repko},
  \citenamefont {Reinhard}, \citenamefont {Nesterenko},\ and\ \citenamefont
  {Kvasil}}]{repko2013}%
  \BibitemOpen
  \bibfield  {author} {\bibinfo {author} {\bibfnamefont {A.}~\bibnamefont
  {Repko}}, \bibinfo {author} {\bibfnamefont {P.-G.}\ \bibnamefont {Reinhard}},
  \bibinfo {author} {\bibfnamefont {V.~O.}\ \bibnamefont {Nesterenko}}, \ and\
  \bibinfo {author} {\bibfnamefont {J.}~\bibnamefont {Kvasil}},\ }\href
  {\doibase 10.1103/PhysRevC.87.024305} {\bibfield  {journal} {\bibinfo
  {journal} {Phys. Rev. C}\ }\textbf {\bibinfo {volume} {87}},\ \bibinfo
  {pages} {024305} (\bibinfo {year} {2013})}\BibitemShut {NoStop}%
\bibitem [{\citenamefont {Alhassid}\ \emph {et~al.}(1982)\citenamefont
  {Alhassid}, \citenamefont {Gai},\ and\ \citenamefont
  {Bertsch}}]{alhassid1982}%
  \BibitemOpen
  \bibfield  {author} {\bibinfo {author} {\bibfnamefont {Y.}~\bibnamefont
  {Alhassid}}, \bibinfo {author} {\bibfnamefont {M.}~\bibnamefont {Gai}}, \
  and\ \bibinfo {author} {\bibfnamefont {G.~F.}\ \bibnamefont {Bertsch}},\
  }\href {\doibase 10.1103/PhysRevLett.49.1482} {\bibfield  {journal} {\bibinfo
   {journal} {Phys. Rev. Lett.}\ }\textbf {\bibinfo {volume} {49}},\ \bibinfo
  {pages} {1482} (\bibinfo {year} {1982})}\BibitemShut {NoStop}%
\bibitem [{\citenamefont {Tamii}\ \emph {et~al.}(2011)\citenamefont {Tamii},
  \citenamefont {Poltoratska}, \citenamefont {von Neumann-Cosel}, \citenamefont
  {Fujita}, \citenamefont {Adachi}, \citenamefont {Bertulani}, \citenamefont
  {Carter}, \citenamefont {Dozono}, \citenamefont {Fujita}, \citenamefont
  {Fujita}, \citenamefont {Hatanaka}, \citenamefont {Ishikawa}, \citenamefont
  {Itoh}, \citenamefont {Kawabata}, \citenamefont {Kalmykov}, \citenamefont
  {Krumbholz}, \citenamefont {Litvinova}, \citenamefont {Matsubara},
  \citenamefont {Nakanishi}, \citenamefont {Neveling}, \citenamefont {Okamura},
  \citenamefont {Ong}, \citenamefont {\"Ozel-Tashenov}, \citenamefont
  {Ponomarev}, \citenamefont {Richter}, \citenamefont {Rubio}, \citenamefont
  {Sakaguchi}, \citenamefont {Sakemi}, \citenamefont {Sasamoto}, \citenamefont
  {Shimbara}, \citenamefont {Shimizu}, \citenamefont {Smit}, \citenamefont
  {Suzuki}, \citenamefont {Tameshige}, \citenamefont {Wambach}, \citenamefont
  {Yamada}, \citenamefont {Yosoi},\ and\ \citenamefont {Zenihiro}}]{tamii2011}%
  \BibitemOpen
  \bibfield  {author} {\bibinfo {author} {\bibfnamefont {A.}~\bibnamefont
  {Tamii}}, \bibinfo {author} {\bibfnamefont {I.}~\bibnamefont {Poltoratska}},
  \bibinfo {author} {\bibfnamefont {P.}~\bibnamefont {von Neumann-Cosel}},
  \bibinfo {author} {\bibfnamefont {Y.}~\bibnamefont {Fujita}}, \bibinfo
  {author} {\bibfnamefont {T.}~\bibnamefont {Adachi}}, \bibinfo {author}
  {\bibfnamefont {C.~A.}\ \bibnamefont {Bertulani}}, \bibinfo {author}
  {\bibfnamefont {J.}~\bibnamefont {Carter}}, \bibinfo {author} {\bibfnamefont
  {M.}~\bibnamefont {Dozono}}, \bibinfo {author} {\bibfnamefont
  {H.}~\bibnamefont {Fujita}}, \bibinfo {author} {\bibfnamefont
  {K.}~\bibnamefont {Fujita}}, \bibinfo {author} {\bibfnamefont
  {K.}~\bibnamefont {Hatanaka}}, \bibinfo {author} {\bibfnamefont
  {D.}~\bibnamefont {Ishikawa}}, \bibinfo {author} {\bibfnamefont
  {M.}~\bibnamefont {Itoh}}, \bibinfo {author} {\bibfnamefont {T.}~\bibnamefont
  {Kawabata}}, \bibinfo {author} {\bibfnamefont {Y.}~\bibnamefont {Kalmykov}},
  \bibinfo {author} {\bibfnamefont {A.~M.}\ \bibnamefont {Krumbholz}}, \bibinfo
  {author} {\bibfnamefont {E.}~\bibnamefont {Litvinova}}, \bibinfo {author}
  {\bibfnamefont {H.}~\bibnamefont {Matsubara}}, \bibinfo {author}
  {\bibfnamefont {K.}~\bibnamefont {Nakanishi}}, \bibinfo {author}
  {\bibfnamefont {R.}~\bibnamefont {Neveling}}, \bibinfo {author}
  {\bibfnamefont {H.}~\bibnamefont {Okamura}}, \bibinfo {author} {\bibfnamefont
  {H.~J.}\ \bibnamefont {Ong}}, \bibinfo {author} {\bibfnamefont
  {B.}~\bibnamefont {\"Ozel-Tashenov}}, \bibinfo {author} {\bibfnamefont
  {V.~Y.}\ \bibnamefont {Ponomarev}}, \bibinfo {author} {\bibfnamefont
  {A.}~\bibnamefont {Richter}}, \bibinfo {author} {\bibfnamefont
  {B.}~\bibnamefont {Rubio}}, \bibinfo {author} {\bibfnamefont
  {H.}~\bibnamefont {Sakaguchi}}, \bibinfo {author} {\bibfnamefont
  {Y.}~\bibnamefont {Sakemi}}, \bibinfo {author} {\bibfnamefont
  {Y.}~\bibnamefont {Sasamoto}}, \bibinfo {author} {\bibfnamefont
  {Y.}~\bibnamefont {Shimbara}}, \bibinfo {author} {\bibfnamefont
  {Y.}~\bibnamefont {Shimizu}}, \bibinfo {author} {\bibfnamefont {F.~D.}\
  \bibnamefont {Smit}}, \bibinfo {author} {\bibfnamefont {T.}~\bibnamefont
  {Suzuki}}, \bibinfo {author} {\bibfnamefont {Y.}~\bibnamefont {Tameshige}},
  \bibinfo {author} {\bibfnamefont {J.}~\bibnamefont {Wambach}}, \bibinfo
  {author} {\bibfnamefont {R.}~\bibnamefont {Yamada}}, \bibinfo {author}
  {\bibfnamefont {M.}~\bibnamefont {Yosoi}}, \ and\ \bibinfo {author}
  {\bibfnamefont {J.}~\bibnamefont {Zenihiro}},\ }\href {\doibase
  10.1103/PhysRevLett.107.062502} {\bibfield  {journal} {\bibinfo  {journal}
  {Phys. Rev. Lett.}\ }\textbf {\bibinfo {volume} {107}},\ \bibinfo {pages}
  {062502} (\bibinfo {year} {2011})}\BibitemShut {NoStop}%
\bibitem [{\citenamefont {Piekarewicz}\ \emph {et~al.}(2012)\citenamefont
  {Piekarewicz}, \citenamefont {Agrawal}, \citenamefont {Col\`o}, \citenamefont
  {Nazarewicz}, \citenamefont {Paar}, \citenamefont {Reinhard}, \citenamefont
  {Roca-Maza},\ and\ \citenamefont {Vretenar}}]{piekarewicz2012}%
  \BibitemOpen
  \bibfield  {author} {\bibinfo {author} {\bibfnamefont {J.}~\bibnamefont
  {Piekarewicz}}, \bibinfo {author} {\bibfnamefont {B.~K.}\ \bibnamefont
  {Agrawal}}, \bibinfo {author} {\bibfnamefont {G.}~\bibnamefont {Col\`o}},
  \bibinfo {author} {\bibfnamefont {W.}~\bibnamefont {Nazarewicz}}, \bibinfo
  {author} {\bibfnamefont {N.}~\bibnamefont {Paar}}, \bibinfo {author}
  {\bibfnamefont {P.-G.}\ \bibnamefont {Reinhard}}, \bibinfo {author}
  {\bibfnamefont {X.}~\bibnamefont {Roca-Maza}}, \ and\ \bibinfo {author}
  {\bibfnamefont {D.}~\bibnamefont {Vretenar}},\ }\href {\doibase
  10.1103/PhysRevC.85.041302} {\bibfield  {journal} {\bibinfo  {journal} {Phys.
  Rev. C}\ }\textbf {\bibinfo {volume} {85}},\ \bibinfo {pages} {041302}
  (\bibinfo {year} {2012})}\BibitemShut {NoStop}%
\bibitem [{\citenamefont {Gazit}\ \emph
  {et~al.}(2006{\natexlab{b}})\citenamefont {Gazit}, \citenamefont {Barnea},
  \citenamefont {Bacca}, \citenamefont {Leidemann},\ and\ \citenamefont
  {Orlandini}}]{tetraedro}%
  \BibitemOpen
  \bibfield  {author} {\bibinfo {author} {\bibfnamefont {D.}~\bibnamefont
  {Gazit}}, \bibinfo {author} {\bibfnamefont {N.}~\bibnamefont {Barnea}},
  \bibinfo {author} {\bibfnamefont {S.}~\bibnamefont {Bacca}}, \bibinfo
  {author} {\bibfnamefont {W.}~\bibnamefont {Leidemann}}, \ and\ \bibinfo
  {author} {\bibfnamefont {G.}~\bibnamefont {Orlandini}},\ }\href {\doibase
  10.1103/PhysRevC.74.061001} {\bibfield  {journal} {\bibinfo  {journal} {Phys.
  Rev. C}\ }\textbf {\bibinfo {volume} {74}},\ \bibinfo {pages} {061001}
  (\bibinfo {year} {2006}{\natexlab{b}})}\BibitemShut {NoStop}%
\bibitem [{\citenamefont {Goerke}\ \emph {et~al.}(2012)\citenamefont {Goerke},
  \citenamefont {Bacca},\ and\ \citenamefont {Barnea}}]{goerke2012}%
  \BibitemOpen
  \bibfield  {author} {\bibinfo {author} {\bibfnamefont {R.}~\bibnamefont
  {Goerke}}, \bibinfo {author} {\bibfnamefont {S.}~\bibnamefont {Bacca}}, \
  and\ \bibinfo {author} {\bibfnamefont {N.}~\bibnamefont {Barnea}},\ }\href
  {\doibase 10.1103/PhysRevC.86.064316} {\bibfield  {journal} {\bibinfo
  {journal} {Phys. Rev. C}\ }\textbf {\bibinfo {volume} {86}},\ \bibinfo
  {pages} {064316} (\bibinfo {year} {2012})}\BibitemShut {NoStop}%
\bibitem [{\citenamefont {Miorelli}\ \emph {et~al.}(2014)\citenamefont
  {Miorelli}, \citenamefont {Bacca}, \citenamefont {Barnea}, \citenamefont
  {Hagen}, \citenamefont {Orlandini},\ and\ \citenamefont
  {Papenbrock}}]{Mirko}%
  \BibitemOpen
  \bibfield  {author} {\bibinfo {author} {\bibfnamefont {M.}~\bibnamefont
  {Miorelli}}, \bibinfo {author} {\bibfnamefont {S.}~\bibnamefont {Bacca}},
  \bibinfo {author} {\bibfnamefont {N.}~\bibnamefont {Barnea}}, \bibinfo
  {author} {\bibfnamefont {G.}~\bibnamefont {Hagen}}, \bibinfo {author}
  {\bibfnamefont {G.}~\bibnamefont {Orlandini}}, \ and\ \bibinfo {author}
  {\bibfnamefont {T.}~\bibnamefont {Papenbrock}},\ }\href@noop {} {\bibfield
  {journal} {\bibinfo  {journal} {in preparation}\ } (\bibinfo {year}
  {2014})}\BibitemShut {NoStop}%
\bibitem [{\citenamefont {Nevo~Dinur}\ \emph {et~al.}(2014)\citenamefont
  {Nevo~Dinur}, \citenamefont {Barnea}, \citenamefont {Ji},\ and\ \citenamefont
  {Bacca}}]{LanczosSumRules}%
  \BibitemOpen
  \bibfield  {author} {\bibinfo {author} {\bibfnamefont {N.}~\bibnamefont
  {Nevo~Dinur}}, \bibinfo {author} {\bibfnamefont {N.}~\bibnamefont {Barnea}},
  \bibinfo {author} {\bibfnamefont {C.}~\bibnamefont {Ji}}, \ and\ \bibinfo
  {author} {\bibfnamefont {S.}~\bibnamefont {Bacca}},\ }\href {\doibase
  10.1103/PhysRevC.89.064317} {\bibfield  {journal} {\bibinfo  {journal} {Phys.
  Rev. C}\ }\textbf {\bibinfo {volume} {89}},\ \bibinfo {pages} {064317}
  (\bibinfo {year} {2014})}\BibitemShut {NoStop}%
\bibitem [{\citenamefont {Angeli}\ and\ \citenamefont
  {Marinova}(2013)}]{angeli2013}%
  \BibitemOpen
  \bibfield  {author} {\bibinfo {author} {\bibfnamefont {I.}~\bibnamefont
  {Angeli}}\ and\ \bibinfo {author} {\bibfnamefont {K.}~\bibnamefont
  {Marinova}},\ }\href {\doibase http://dx.doi.org/10.1016/j.adt.2011.12.006}
  {\bibfield  {journal} {\bibinfo  {journal} {Atomic Data and Nuclear Data
  Tables}\ }\textbf {\bibinfo {volume} {99}},\ \bibinfo {pages} {69 } (\bibinfo
  {year} {2013})}\BibitemShut {NoStop}%
\end{thebibliography}%
\bibliographystyle{apsrev4-1}

\end{document}